\documentclass[twocolumn,english,aps,prb,amsmath,showpacs,amssymb]{revtex4}
\usepackage[T1]{fontenc}
\usepackage[latin9]{inputenc}
\usepackage{bm}
\usepackage{amsmath}
\usepackage{graphicx}
\usepackage{amssymb}
\usepackage{esint}

\makeatletter

\providecommand{\tabularnewline}{\\}

\@ifundefined{textcolor}{}
{%
 \definecolor{BLACK}{gray}{0}
 \definecolor{WHITE}{gray}{1}
 \definecolor{RED}{rgb}{1,0,0}
 \definecolor{GREEN}{rgb}{0,1,0}
 \definecolor{BLUE}{rgb}{0,0,1}
 \definecolor{CYAN}{cmyk}{1,0,0,0}
 \definecolor{MAGENTA}{cmyk}{0,1,0,0}
 \definecolor{YELLOW}{cmyk}{0,0,1,0}
 }

\usepackage{bm}

\makeatletter

\providecommand{\tabularnewline}{\\}

\@ifundefined{textcolor}{}{\definecolor{BLACK}{gray}{0}\definecolor{WHITE}{gray}{1}\definecolor{RED}{rgb}{1,0,0}\definecolor{GREEN}{rgb}{0,1,0}\definecolor{BLUE}{rgb}{0,0,1}\definecolor{CYAN}{cmyk}{1,0,0,0}\definecolor{MAGENTA}{cmyk}{0,1,0,0}\definecolor{YELLOW}{cmyk}{0,0,1,0}}

\usepackage{bm}

\makeatletter

\providecommand{\tabularnewline}{\\}

\@ifundefined{textcolor}{}{\definecolor{BLACK}{gray}{0}\definecolor{WHITE}{gray}{1}\definecolor{RED}{rgb}{1,0,0}\definecolor{GREEN}{rgb}{0,1,0}\definecolor{BLUE}{rgb}{0,0,1}\definecolor{CYAN}{cmyk}{1,0,0,0}\definecolor{MAGENTA}{cmyk}{0,1,0,0}\definecolor{YELLOW}{cmyk}{0,0,1,0}}

\newcommand{\e}{\epsilon}

\def\lapp{\lower.35em\hbox{$\stackrel{\textstyle<}{\sim}$}}

\newlength{\textwidthm}

\makeatother

\usepackage{babel}

\makeatother

\usepackage{babel}

\makeatother

\usepackage{babel}

\makeatother

\usepackage{babel}

\makeatother

\usepackage{babel}

\makeatother

\usepackage{babel}

\makeatother

\usepackage{babel}

\makeatother

\usepackage{babel}

\begin{document}


\title{Unified description of the dc-conductivity of monolayer and bilayer
graphene at finite densities based on resonant scatterers}

\author{Aires Ferreira $^{1}$, J. Viana-Gomes$^{1}$, Johan Nilsson$^{2}$,
E. R. Mucciolo$^{3}$, N. M. R. Peres$^{1}$, and A. H. Castro Neto$^{4}$}

\affiliation{$^{1}$ Department of Physics and Center of Physics, University of
Minho, P-4710-057, Braga, Portugal}

\affiliation{$^{2}$ Department of Physics, University of Gothenburg, 412 96 Gothenburg,
Sweden}

\affiliation{$^{3}$ Department of Physics, University of Central Florida, Orlando,
Florida 32816, USA}

\affiliation{$^{4}$ Department of Physics, Boston University, 590 Commonwealth
Avenue, Boston, Massachusetts 02215, USA}

\date{\today}
\begin{abstract}
We show that a coherent picture of the dc conductivity of monolayer
and bilayer graphene at finite electronic densities emerges upon considering
that strong short-range potentials are the main source of scattering
in these two systems. The origin of the strong short-range potentials
may lie in adsorbed hydrocarbons at the surface of graphene. The equivalence
among results based on the partial-wave description of scattering,
the Lippmann-Schwinger equation, and the $T$ matrix approach is established.
Scattering due to resonant impurities close to the neutrality point
is investigated via a numerical computation of the Kubo formula using
a kernel polynomial method. We find that relevant adsorbate species
originate impurity bands in monolayer and bilayer graphene close to
the Dirac point. In the midgap region, a plateau of minimum conductivity
of about $e^{2}/h$ (per layer) is induced by the resonant disorder.
In bilayer graphene, a large adsorbate concentration can develop an
energy gap between midgap and high-energy states. As a consequence,
the conductivity plateau is supressed near the edges and a {}``conductivity
gap\textquotedblright{} takes place. Finally, a scattering formalism
for electrons in biased bilayer graphene, taking into account the
degeneracy of the spectrum, is developed and the dc conductivity of
that system is studied. 
\end{abstract}
\maketitle

\section{Introduction\label{introd}}

In his famous book,\cite{peierls} Peierls noted that in three dimensions
the first Born approximation (FBA) cannot be used to deal with short-range
potentials in general, even when the potential is not too strong.
The reason lies in the fact that the FBA overestimates the value of
the scattering cross section and modifies the energy dependence of
the latter relative to the exact result. The fundamental reason why
this effect takes place has its roots in the modification of the wave
function within the region where the potential is finite. There, even
for moderate potentials, the wave function is strongly deformed relative
to the plane wave used in the FBA.

Since his main concern was nuclear physics, Peierls did not address
the validity of the FBA in systems of reduced dimensions. Contrary
to nuclear physics, some condensed matter systems impose dimensional
constraints on the electronic motion --- a direct consequence of the
lattice structure of the given solid. Electrons moving in graphene
face the most dramatic dimensional constraint, being forced to move
along a strictly two-dimensional plane formed by a honeycomb lattice
of carbon atoms.\cite{nov04,pnas,rmp,PeresRMP2010} In bilayer graphene,
electrons are also confined to move in two dimensions. Since bilayer
systems are a stacking of two graphene sheets, the electrons may,
additionally, hop between the layers.

Scattering cross sections in condensed matter physics are of ultimate
importance for the interpretation of dc transport in solids, especially
concerning the effect of localized impurities. These can be described
by either short-range or long-range potentials. Following Peierls,\cite{peierls}
the correct interpretation of the conductivity of a metal at low temperatures
may require a description of electronic scattering by impurities beyond
the FBA: this is particularly true if the impurities give rise to
strong short-range potentials.

In systems such as monolayer and bilayer graphene, where the electronic
density can be tuned between 0 and $\sim10^{14}$ cm$^{-2}$, \cite{efetov_kim_14}
computing the correct dependence of the cross section on the Fermi
energy is a crucial ingredient for a meaningful interpretation of
the data. Since the early days of graphene physics,\cite{nov04,pnas}
it became clear that the conductivity of monolayer graphene shows
slightly sublinear dependence on electronic density. On the other
hand, the conductivity of bilayer graphene shows, consistently, a
robust linear dependence on the backgate potential. Both monolayer
and bilayer graphene-based field-effect devices use sheets from flakes
produced in exactly the same manner, i.e., via exfoliation of graphite.
(More recently, graphene has been isolated via epitaxial growth on
SiC \cite{epitaxial_SiC} and chemical vapor depositions on metal
surfaces.\cite{CVD_copper,CVD_Niquel}) It is now believed that the
main sources of electronic scattering in exfoliated graphene are introduced
during the device fabrication process.

The sources of disorder in graphene can vary. They can be due to adsorbed
chemical species, such as hydrogen atoms or hydrocarbon molecules,
random strain,\cite{vitorstrain} rippling \cite{meyerGeimsuspended,kastripples}
and scrolling,\cite{foglerscroll} and electrostatic random potentials
at the surface of the silicon oxide substrate caused by charged impurities.\cite{Crommie2009,Sarma1,Sarma2,Sarma3}
(Chemically synthesized graphene displays alternative scattering mechanisms.\cite{lines_of_charge})

It is widely accepted that the strong carrier density fluctuations
(electron-hole puddles) observed close to the neutrality or Dirac
point are due to localized subsurface charged impurities.\cite{Crommie2009,yacoby}
Whether charged impurities are also the limiting source of scattering
in \textit{doped} graphene (i.e. away from the neutrality point) remains
unclear. In addition to charged scatterers, resonant scattering due
to adsorbed hydrocarbons \cite{wehlingII} is currently ascending
in the list of candidates limiting the electronic mobility in graphene.\cite{Irradiation_Ions_SLG,miguelmonteverde,Dpeak,Katoch2010}
As we show in Sec. \ref{sec_resonant}, adsorbed hydrocarbons can
effectively act as strong short-range scatterers. Strong, short-range,
resonant scatterers can be mimicked by vacancies in a lattice model.\cite{nmrPRB06,vitorpaco,vitordisorder}
In magneto-optical transport studies of graphene, short-range scattering
seems essential to explain the width of the cyclotron peak at high
magnetic field. \cite{cyclotron_peak}

Since the sources of scattering are likely introduced during the fabrication
process, they must be the same for both monolayer and bilayer graphene.
Therefore, a consistent theoretical description of the conductivity
of graphene, at low temperatures and finite electronic densities,
must be able to describe the experimental curves of both monolayer
and bilayer graphene by invoking the same source of scattering. In
this paper, we show that such a consistent theoretical description
can be achieved by considering strong short-range potentials whose
origin may lie in adsorbed chemical species at the surface of the
material. Instrumental to our description is the critical analysis
developed by Peierls: Calculation of the exact scattering cross sections
is essential for a correct interpretation of the experimental data.

Before studying the dc conductivity for both monolayer and bilayer
graphene at finite electronic densities, a task we defer to Sec. \ref{DCcond},
we first survey the scattering theory for electrons in these systems
in Sec. \ref{partialwave}. This first step is essential for comprehension
of the remaining sections.

In Sec. \ref{DCcond}, we show, using a simple and intuitive model,
that the effect of adsorbed chemical species on graphene is equivalent
to that of very strong on-site short-range potentials --- the so-called
resonant scatterers. Here, we use lattice-based numerical calculations
of the density of states to show in some detail how this class of
impurities affects the electronic structure of monolayer and bilayer
graphene. Using a continuous formulation, we also show that the semiclassical
dc conductivity of both monolayer and bilayer graphene at finite densities
can be easily calculated using the intuitive approach to scattering
given the partial-wave analysis. We apply the developed formalism
to resonant scatterers, and show that this type of short-range disorder
accounts well for experimental data.

Further, we demonstrate the need for the computation of exact electronic
scattering amplitudes when applying the Boltzmann approach to strong
short-range potentials, an issue overlooked in the literature that
we re-examine here. The validity of the semiclassical results at finite
electronic densities and low impurity densities is established via
a $T$-matrix calculation of the Kubo dc conductivity. Finally, by
means of a numerical calculation based on the kernel polynomial method
(KPM), we illustrate the breakdown of the semiclassical picture for
electronic densities close to the neutrality point. These simulations
explore the limit of finite impurity density, thus fully taking into
account interference effects neglected in the Boltzmann approach.

In Sec. \ref{sec_scattBBL}, we adapt the formalism of Secs. \ref{partialwave}
and \ref{DCcond} to describe scattering when a perpendicular electric
field is applied to bilayer graphene. Conclusions are drawn in Sec.
\ref{sec_conclusion}. Several technical aspects of our results are
given in the Appendix.

We note that transport in monolayer and bilayer graphene was addressed
by some of us in an ealier publication.\cite{nilsson2} However, it
is important to remark that in the present work our goal is to provide
a unified description of transport in both systems based on the same
scattering mechanism. Also, it is shown that the transport properties
of the bilayer graphene can be understood in a much simpler, intuitive,
and transparent way using the \textit{standard} scattering formalism
of partial waves. In this regard, our present work is complementary
to the study developed in Ref. \onlinecite{nilsson2}. That is,
the present work closes the circle of showing that for both graphene
and its bilayer, a coherent and unified description of dc transport
in these systems can be described by one and the same formalism, be
it the more formal and mathematically demanding one of the transfer
matrix or the intuitive and simple one of partial waves.

\section{Partial-wave analysis in graphene\label{partialwave}}

As discussed in Sec.~\ref{introd}, calculation of the dc conductivity
of a metal requires computing transport cross section as accurately
as possible. A well-established approach is based on the computation
of the phase shifts induced in the scattered electron wave function
by the scattering potential. If the phase shifts are known exactly,
so is the cross section. Below, we set the notation and introduce
the central quantities needed in this work by giving a concise presentation
of the phase-shift approach to scattering in the context of graphene
and its bilayer. \cite{coulombnovikov,katsnelsonBL,basko,hentschel}
These results are later used in Sec. \ref{DCcond}. Also, and to the
best of our knowledge, the scattering theory for electrons in a biased
graphene bilayer has not been developed so far in the literature,
and therefore it is presented in Sec. \ref{sec_scattBBL}.

Scattering theory states that the large-distance wave function of
a particle in the presence of a scattering potential (with cylindrical
symmetry) must have the form (in two dimensions) \begin{equation}
\psi(\mathbf{r})\simeq e^{ik_{i}x}+f(\theta)\frac{e^{ik_{f}r}}{\sqrt{r}}\,,\label{eq_scatt2D}\end{equation}
 where $\mathbf{k}_{i}=(k_{i},0)$ and $\mathbf{k}_{f}=k_{f}(\cos\theta,\sin\theta)$
are the momentum of the incoming and scattered waves, respectively;
clearly, for elastic scattering, we have $k_{i}=k_{f}=k$. The scattering
amplitude $f(\theta)$ can be written in terms of the phase shifts
$\delta_{m}$ associated with the partial-wave expansion of the scattered
wave function in the basis of angular momentum states. In Eq. (\ref{eq_scatt2D}),
the first term represents the incoming particle, with the incoming
momentum oriented along the $x$ axis, and the second one represents
the cylindrical scattered wave function.

As it stands, Eq. (\ref{eq_scatt2D}) holds for the two-dimensional
Schrödinger equation.\cite{2Dscatt} However, for both monolayer and
bilayer graphene, the large distance behavior of the wave function
differs slightly, but significantly, from Eq. (\ref{eq_scatt2D}).

\subsection{Electronic scattering in graphene}

\begin{figure}[ht]
 \centering{} \includegraphics[width=8cm]{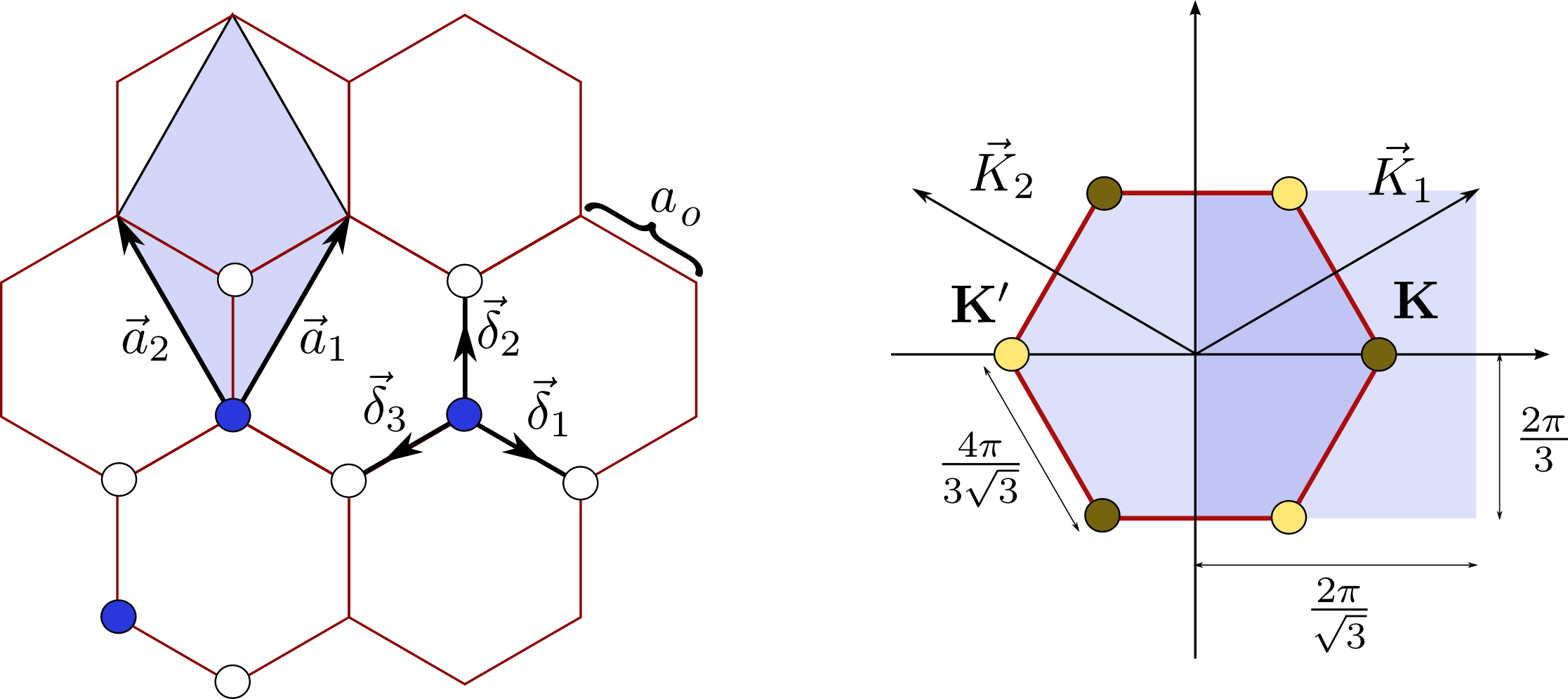} \caption{\label{fig_monolayer_lattice} (Color online) Lattice structure and
Brillouin zone of monolayer graphene. \textbf{Left:} Hexagonal lattice
of graphene, with the next nearest neighbor, $\bm{\delta}_{i}$, and
the primitive, $\bm{a}_{i}$, vectors depicted. The area of the primitive
cell is $A_{c}=3\sqrt{3}a_{0}^{2}/2\simeq5.1$ \AA{}$^{2}$, and $a_{0}\simeq1.4$
\AA{}. \textbf{Right:} Brillouin zone of graphene, with the Dirac
points $\bm{K}$ and $\bm{K}'$ indicated. Close to these points,
the dispersion of graphene is conical and the density of states is
proportional to the absolute value of the energy.}

\end{figure}

For graphene, the motion of the electrons in the $\pi$ orbitals is,
at low energies, described by the two-dimensional massless Dirac Hamiltonian,
reading \cite{rmp} \begin{equation}
H_{\bm{K}}=v_{F}\bm{\sigma}\cdot\mathbf{p}\,,\label{eq_dirac}\end{equation}
 where the Fermi velocity is defined as $v_{F}=3ta_{0}/(2\hbar)$,
$t$ is the hopping integral between the $p_{z}$ orbitals of two
adjacent carbon atoms, and $a_{0}\approx1.4$ \AA{}\, is the carbon-carbon
distance in graphene (see Fig. \ref{fig_monolayer_lattice}). The
vector $\bm{\sigma}$ is written in terms of Pauli's matrices as $\bm{\sigma}=(\sigma_{x},\sigma_{y})$,
and $\mathbf{p}$ is the momentum operator. The vector $\bm{K}$ denotes
one of the two (inequivalent) edge points of the Brillouin zone, also
called Dirac points or valleys. Because neutral graphene is half-filled
(i.e., the $\pi$ orbitals contain one electron), these two points
control the low-energy physics. Depending on the nature of disorder
and the Fermi energy, coupling between momentum states from different
valleys can take place. Intervalley scattering is known to induce
weak localization corrections to the conductivity and, ultimately,
fully localize states in the thermodynamic limit at zero temperature.
\cite{McCann2006,Tikhonenko,EduardoReview}

In what follows, we assume that the two Dirac points, $\bm{K}$ and
$\bm{K}'$, can be treated independently. This procedure is justified
because intervalley scattering (known to occur for short-range scatterers)
manifests itself primarily in the coherent regime, through backscattering
interference. For low concentrations of scattering centers, finite
sample size, and finite temperatures (the typical experimental situation),
coupling between the Dirac points can be neglected when considering
high enough electronic densities. Hence, with the exception of the
lattice calculations (Secs. \ref{sub:tmatrix} and \ref{sub:KPM}),
we neglect intervalley scattering in the continuous model calculations
and introduce the valley degeneracy index, $g_{v}=2$. In Cartesian
coordinates, the eigenstate of the Hamiltonian in Eq.~(\ref{eq_dirac})
has the explicit form \begin{equation}
\Psi_{\pm}(\mathbf{r})=\frac{1}{\sqrt{2A}}\left[\begin{array}{c}
1\\
\pm e^{i\theta_{\mathbf{k}}}\end{array}\right]e^{i\mathbf{k}\cdot\mathbf{r}}\,,\label{eq_spinora}\end{equation}
 with $\theta_{\mathbf{k}}=\arctan(k_{y}/k_{x})$ and $A$ denoting
the total area of the system. The energy eigenvalues corresponding
to the eigenfunction in Eq.~(\ref{eq_spinora}) are $E=\pm v_{F}\hbar k$.
From the latter follows the density of states per spin and per unit
cell, $\rho(E)=2\vert E\vert/(\pi\sqrt{3}t^{2})$, where the contribution
from the two valleys has been taken into account. The probability
density current reads \cite{Schif} \begin{equation}
\bm{J}=v_{F}\Psi_{\pm}^{\dagger}\boldsymbol{\sigma}\Psi_{\pm}\,.\end{equation}
 For the study of scattering, it is more convenient to recast the
Hamiltonian in Eq.~(\ref{eq_dirac}) in cylindrical coordinates $r$
and $\theta$ as \begin{equation}
H_{\bm{K}}=-iv_{F}\hbar\left[\begin{array}{cc}
0 & \hat{L}_{-}\\
\hat{L}_{+} & 0\end{array}\right]\,,\label{eq_diraccylind}\end{equation}
 where the operators $\hat{L}_{\pm}=e^{\pm i\theta}(\partial_{r}\pm ir^{-1}\partial_{\theta})$
act as rising/lowering operators, according to the following result
\begin{equation}
\hat{L}_{\pm}[C_{m}(kr)e^{i\theta m}]=\mp kC_{m\pm1}(kr)e^{i\theta(m\pm1)}\,.\label{eq_ladder}\end{equation}
 In Eq.~(\ref{eq_ladder}), the function $C_{m}(kr)$ stands for
$J_{m}(kr)$ and $Y_{m}(kr)$, the first-kind and second-kind Bessel
functions, respectively, and for the Hankel functions of the first
kind $H_{m}^{(1)}$ and second kind $H_{m}^{(2)}$. For the modified
Bessel function $K_{m}(kr)$ we have \begin{equation}
\hat{L}_{\pm}[K_{m}(kr)e^{i\theta m}]=-kK_{m\pm1}(kr)e^{i\theta(m\pm1)}\,.\label{eq_ladder_modified}\end{equation}
 In cylindric coordinates, the radial probability density current
reads \begin{equation}
J_{r}=v_{F}\Psi_{\pm}^{\dagger}\sigma_{r}\Psi_{\pm}\,,\label{eq_JrDirac}\end{equation}
 where $\sigma_{r}$ is defined as \begin{equation}
\sigma_{r}=\left[\begin{array}{cc}
0 & e^{-i\theta}\\
e^{i\theta} & 0\end{array}\right]\,.\end{equation}
 The tangential component of the probability density current reads
$J_{\theta}=v_{F}\Psi_{\pm}^{\dagger}\sigma_{\theta}\Psi_{\pm}$,
with $\sigma_{\theta}=\sigma_{r}\textrm{diag}(i,-i)$, and where $\textrm{diag}(i,-i)$
represents a diagonal matrix. Let us now derive, for massless Dirac
electrons in two dimensions, the equivalent of the asymptotic wave
function in Eq.~(\ref{eq_scatt2D}). To that end, we note that a
state having the form \begin{equation}
\Psi_{m}(r,\theta)=\frac{1}{\sqrt{2A}}\left[\begin{array}{c}
J_{m}(kr)\\
\pm ie^{i\theta}J_{m+1}(kr)\end{array}\right]e^{im\theta}\,\label{eq:solution_dirac_slg}\end{equation}
 is also an eigenstate of the Hamiltonian in Eq.~(\ref{eq_diraccylind}).
We start by assuming that the asymptotic (large $r$) behavior of
the wave function in the angular momentum channel $m$ has the form
(from here on, we consider only $E>0$) \begin{eqnarray}
\Psi_{m}(r,\theta) & \simeq & \sqrt{\frac{1}{\pi Akr}}\left[\begin{array}{c}
\cos(kr-\lambda_{m}+\delta_{m})\\
ie^{i\theta}\sin(kr-\lambda_{m}+\delta_{m})\end{array}\right]\nonumber \\
 &  & \times\, e^{i(m\theta+\delta_{m})}\,,\label{eq_psi_dirac_angular}\end{eqnarray}
 an ansatz inspired by the fact that the Dirac equation for graphene
is a set of two coupled first-order differential equations and in
the asymptotic limit of the Bessel functions at large $r$:\cite{abramowitz}
\begin{eqnarray}
J_{m}(x)=\sqrt{\frac{2}{\pi x}}\cos(x-\lambda_{m})\,,\label{eq_BesselJ_xlarge}\\
Y_{m}(x)=\sqrt{\frac{2}{\pi x}}\sin(x-\lambda_{m})\,,\label{eq_BesselY_xlarge}\end{eqnarray}
 with $\lambda_{m}=m\pi/2+\pi/4$. Using Eq.~(\ref{eq_psi_dirac_angular}),
we write the total wave function as an expansion in partial waves,
reading \begin{equation}
\Psi(r,\theta)=\sum_{m=-\infty}^{\infty}i^{m}\Psi_{m}(r,\theta)\,.\end{equation}
 Exploiting of the relation \begin{equation}
e^{ikr\cos\theta}=\sum_{m=-\infty}^{\infty}i^{m}e^{im\theta}J_{m}(kr)\,,\end{equation}
 we obtain \begin{equation}
\Psi(\mathbf{r})\simeq\frac{1}{\sqrt{2A}}\left[\begin{array}{c}
1\\
1\end{array}\right]e^{ikx}+\frac{1}{\sqrt{2A}}\left[\begin{array}{c}
1\\
e^{i\theta}\end{array}\right]f(\theta)\frac{e^{ikr}}{\sqrt{r}}\,,\label{eq_scattDirac}\end{equation}
 with the scattering amplitude reading \begin{equation}
f(\theta)=\sqrt{\frac{2i}{\pi k}}\sum_{m=-\infty}^{\infty}e^{im\theta}e^{i\delta_{m}}\sin\delta_{m}\,.\label{eq_scatt_amplitude}\end{equation}
 It is a simple exercise to show that the first term in Eq.~(\ref{eq_scattDirac})
corresponds to a flux $J_{x}=v_{F}/A$ (and $J_{y}=0$), whereas the
second term corresponds to a radial flux $J_{r}=v_{F}\vert f(\theta)\vert^{2}/(rA)$
(and $J_{\theta}=0$). Thus, according to the usual definition of
the differential cross section, $\sigma(\theta)$, it follows that
\begin{equation}
\sigma(\theta)=\vert f(\theta)\vert^{2}\,.\label{eq_diffcrosssection}\end{equation}

Before we turn to scattering in bilayer graphene, it will be useful,
for later use, to introduce other asymptotic forms of the Bessel functions
$J_{m}(x)$, $Y_{m}(x)$, and $K_{m}(x)$, in addition to those already
given in Eqs.~(\ref{eq_BesselJ_xlarge}) and (\ref{eq_BesselY_xlarge}).
For large $x$, we have \cite{abramowitz} \begin{equation}
K_{m}(x)=\sqrt{\frac{\pi}{2x}}e^{-x}\,.\label{eq_Km_largex}\end{equation}
 For $x\ll1$, the asymptotic forms read \cite{abramowitz} \begin{equation}
J_{m}(x)=(x/2)^{m}\Gamma^{-1}(m+1)\,,\label{eq_Jm_smallx}\end{equation}
 \begin{eqnarray}
Y_{0}(x) & = & 2\pi^{-1}\ln x\,,\label{eq_Y0_smallx}\\
\nonumber \\Y_{m}(x) & = & -\pi^{-1}\Gamma(m)(x/2)^{-m}\,,\quad m=1,2,\ldots\,,\label{eq_Ym_smallx}\end{eqnarray}
 and \begin{eqnarray}
K_{0}(x) & = & -\ln x\,,\label{eq_K0_smallx}\\
\nonumber \\K_{m}(x) & = & 2^{-1}\Gamma(m)(x/2)^{-m}\,,\quad m=1,2,\ldots\,,\label{eq_Km_smallx}\end{eqnarray}
 where $\Gamma(x)$ is the gamma function. We now consider scattering
in bilayer graphene.

\subsection{Electronic scattering in bilayer graphene\label{subsecBL}}

\begin{figure}[ht]
 \centering{}\includegraphics[width=8cm]{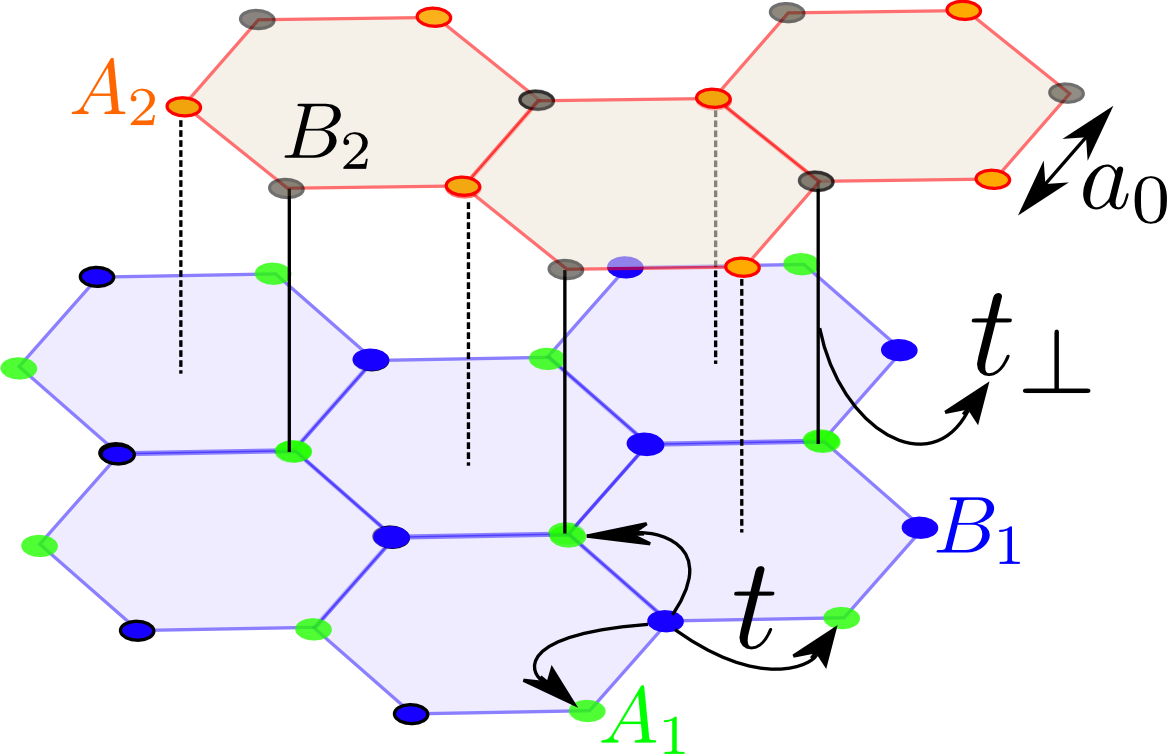} \caption{\label{fig_bilayer_lattice} (Color online) Lattice structure of bilayer
graphene. The atoms labeled $A_{1}$ and $B_{1}$ lie on the bottom
graphene layer, whereas atoms $A_{2}$ and $B_{2}$ are in the top
layer. Electrons can hop between layers via a perpendicular hopping
parameter $t_{\perp}$ between the carbon atom $A_{1}$ and carbon
atom $B_{2}$ (connected by solid lines). The Brillouin zone of bilayer
graphene is the same as that of monolayer graphene (see Fig.~\ref{fig_monolayer_lattice}).}

\end{figure}

Bilayer graphene has four atoms per unit cell, with the two honeycomb
sheets arranged according to a Bernal stacking, as shown in Fig.~\ref{fig_bilayer_lattice}.
Two of the atoms belonging to each of the layers are on top of each
other (atoms $A_{1}$ and $B_{2}$, in Fig.~\ref{fig_bilayer_lattice}),
allowing for interlayer hopping. This process is represented by a
hopping parameter, $t_{\perp}\approx0.5$ eV.\cite{castro,castro_2010}
The other two carbon atoms, labeled $A_{2}$ and $B_{1}$ in Fig.~\ref{fig_bilayer_lattice},
are not coupled to the carbon atoms of the other layer, in accordance
with the assumptions of the minimal model for electronic motion in
bilayer graphene.

The band structure of bilayer graphene has four bands, but the low-energy
physics ($\vert E\vert\ll t_{\perp}$) can be described by an effective
model of only two bands,\cite{McCann1,castro,castro_2010} where the
atoms linked by $t_{\perp}$ are projected out since they describe
high-energy bands: the dimmer of atoms $A_{1}$ and $B_{2}$, linked
by $t_{\perp}$, form a two-level system with energy levels $\pm t_{\perp}$.
Additionally, the atoms in the two sheets can be made nonequivalent
by applying an electric field perpendicular to the sheets, in this
way inducing a gap in the spectrum (the electrostatic potential difference
between the two layers is $2V$).\cite{McCann1,McCann2,castro,castro_2010}

The derivation of the effective Hamiltonian is straightforward. We
write the full Hamiltonian as \begin{equation}
H=\left[\begin{array}{cccc}
V & 0 & 0 & \hat{\pi}\\
0 & -V & \hat{\pi}^{\dag} & 0\\
0 & \hat{\pi} & -V & -t_{\perp}\\
\hat{\pi}^{\dag} & 0 & -t_{\perp} & V\end{array}\right]\equiv\left[\begin{array}{cc}
H_{\textrm{L}} & H_{\textrm{LH}}\\
H_{\textrm{LH}}^{\dag} & H_{\textrm{H}}\end{array}\right]\,\,,\end{equation}
 where the columns in the Hamiltonian are labeled by the four atoms
in the unit cell. In ascending order, this labeling is $B_{1}$, $A_{2}$,
$B_{2}$, and $A_{1}$. The operator $\hat{\pi}$ is defined as $\hat{\pi}\equiv v_{F}\left(\hat{p}_{x}+i\hat{p}_{y}\right)$.
The eigenproblem $H\vert\psi\rangle=E\vert\psi\rangle$ can be written
as \begin{equation}
\left[\begin{array}{cc}
H_{\textrm{L}} & H_{\textrm{LH}}\\
H_{\textrm{LH}}^{\dag} & H_{\textrm{H}}\end{array}\right]\left[\begin{array}{c}
\vert\varphi\rangle\\
\vert\chi\rangle\end{array}\right]=E\left[\begin{array}{c}
\vert\varphi\rangle\\
\vert\chi\rangle\end{array}\right]\,.\label{eq_HBLfull}\end{equation}
 It follows from Eq.~(\ref{eq_HBLfull}) that \begin{equation}
H_{\textrm{L}}\vert\varphi\rangle+H_{\textrm{LH}}(E-H_{\textrm{H}})^{-1}H_{\textrm{LH}}^{\dag}\vert\varphi\rangle=E\vert\varphi\rangle\,,\end{equation}
 and considering that $t_{\perp}\gg(V,\vert E\vert)$, we have $H_{\textrm{BL}}\vert\varphi\rangle=E\vert\varphi\rangle$,
with \cite{McCann1} \begin{equation}
H_{\textrm{BL}}=V\sigma_{z}-\frac{V}{t_{\perp}^{2}}\left[\begin{array}{cc}
\hat{\pi}\hat{\pi}^{\dag} & 0\\
0 & -\hat{\pi}^{\dag}\hat{\pi}\end{array}\right]+\frac{1}{t_{\perp}}\left[\begin{array}{cc}
0 & (\hat{\pi})^{2}\\
(\hat{\pi}^{\dag})^{2} & 0\end{array}\right]\,.\label{eq_HBLVneZ}\end{equation}
 To keep things simple, in what follows we consider the case $V=0$;
later we discuss the case $V\ne0$. In cylindric coordinates, the
Hamiltonian, Eq.~(\ref{eq_HBLVneZ}), is written as \begin{equation}
H_{\textrm{BL}}=-\frac{v_{F}^{2}\hbar^{2}}{t_{\perp}}\left[\begin{array}{cc}
0 & \hat{L}_{-}^{2}\\
\hat{L}_{+}^{2} & 0\end{array}\right]\,,\label{eq_BL2comp}\end{equation}
 and the eigenfunctions (regular at the origin) can be written as
\begin{equation}
\Psi_{m}(r,\theta)=\frac{1}{\sqrt{2A}}\left[\begin{array}{c}
J_{m}(kr)\\
\mp e^{2i\theta}J_{m+2}(kr)\end{array}\right]e^{im\theta}\,,\end{equation}
 to which the eigenvalues $E=\pm v_{F}^{2}\hbar^{2}k^{2}/t_{\perp}$
correspond. From the latter result follows the density of states per
spin and per unit cell, $\rho(E)=t_{\perp}/(\pi\sqrt{3}t^{2})$, where
we have included a factor of 2 coming from the valley degeneracy.\onlinecite{comment}

It is important to stress two differences between the Hamiltonians
in Eqs.~(\ref{eq_dirac}) and (\ref{eq_BL2comp}) regarding boundary
conditions and the nature of the scattering states. To be concrete,
let us assume that the electron is subjected to a potential well of
the form $V(r)=V_{0}\theta(R-r)$. In the case of the Dirac Hamiltonian,
the boundary conditions at $r=R$ imply continuity of the two components
of the spinors, whereas for the bilayer Hamiltonian we have to impose
continuity of both the components of the spinors and their first derivative.
The second aspect is related to the fact that elastic scattering conserves
energy. Thus, since in bilayer graphene we have $E=\pm v_{F}^{2}\hbar^{2}k^{2}/t_{\perp}$,
and keeping the energy constant, say $E>0$, as in any scattering
process, there are two admissible solutions: a real solution, $k=\sqrt{t_{\perp}E}/(v_{F}\hbar)$,
and a purely imaginary one, $k=i\sqrt{t_{\perp}E}/(v_{F}\hbar)$.
Therefore, bilayer graphene supports evanescent modes at the interface
$r=R$. This fact is essential to satisfy the boundary conditions
obeyed by the wave function.\cite{klein}

As in the case of the Dirac Hamiltonian, we have to derive the form
of the probability density current for electrons described by the
Hamiltonian in Eq.~(\ref{eq_BL2comp}). The usual procedure \cite{Schif}
gives that any component $J_{\ell}$ of the current has the form \begin{equation}
J_{\ell}=2\frac{v_{F}^{2}\hbar}{t_{\perp}}\textrm{Im}\Psi^{\dagger}\hat{J}_{\ell}\Psi\,,\label{eq_JellBL}\end{equation}
 where for $\ell=x,y$ we have \begin{equation}
\hat{J}_{x}=\sigma_{x}\partial_{x}+\sigma_{y}\partial_{y}\,,\end{equation}
 and \begin{equation}
\hat{J}_{y}=\sigma_{y}\partial_{x}-\sigma_{x}\partial_{y}\,.\end{equation}
 For the radial component, $\ell=r$, we have \begin{equation}
\hat{J}_{r}=\left[\begin{array}{cc}
0 & e^{-2i\theta}(\partial_{r}+ir^{-1}\partial_{\theta})\\
e^{2i\theta}(\partial_{r}-ir^{-1}\partial_{\theta}) & 0\end{array}\right]\,,\end{equation}
 and for the tangential component, $\ell=\theta$, we have \begin{equation}
\hat{J}_{\theta}=\left[\begin{array}{cc}
0 & -ie^{-2i\theta}(\partial_{r}-ir^{-1}\partial_{\theta})\\
ie^{2i\theta}(\partial_{r}+ir^{-1}\partial_{\theta}) & 0\end{array}\right]\,,\end{equation}

Taking into account that the Hamiltonian in Eq.~(\ref{eq_BL2comp})
forms a set of two coupled second-order differential equations, we
assume that the asymptotic (large $r$) behavior of the wave function
in the angular momentum channel $m$ has the form \begin{eqnarray}
\Psi_{m}(r,\theta) & \simeq & \sqrt{\frac{1}{\pi Akr}}\left[\begin{array}{c}
\cos(kr-\lambda_{m}+\delta_{m})\\
e^{2i\theta}\cos(kr-\lambda_{m}+\delta_{m})\end{array}\right]\nonumber \\
 &  & \times e^{i(m\theta+\delta_{m})}\,.\label{eq_psi_BL_angular}\end{eqnarray}
 Following the same procedure used to derive Eq.~(\ref{eq_scattDirac}),
we can show that the large-$r$ behavior of the total electronic wave
function in graphene bilayer in the presence of a potential has the
form \begin{equation}
\Psi(\mathbf{r})\simeq\frac{1}{\sqrt{2A}}\left(\begin{array}{c}
1\\
1\end{array}\right)e^{ikx}+\frac{1}{\sqrt{2A}}\left(\begin{array}{c}
1\\
e^{2i\theta}\end{array}\right)f(\theta)\frac{e^{ikr}}{\sqrt{r}}\,.\label{eq_scattBL}\end{equation}
 Using Eq.~(\ref{eq_JellBL}), we can easily conclude that the first
term in Eq.~(\ref{eq_scattBL}) corresponds to a flux $J_{x}=2v_{F}^{2}\hbar k/(At_{\perp})\equiv v/A$,
where $v$ is the velocity of the particle, and that the second term
corresponds to a radial flux of the form $J_{r}=2v_{F}^{2}\hbar k\vert f(\theta)\vert^{2}/(rAt_{\perp})\equiv v\vert f(\theta)\vert^{2}/(Ar)$,
with $f(\theta)$ still given by Eq.~(\ref{eq_scatt_amplitude}).
As before, it follows that the differential cross section is given
by Eq.~(\ref{eq_diffcrosssection}).

In Sec.~\ref{DCcond}, we apply the this formalism to the case of
a potential well described by the potential $V(r)=V_{0}\theta(R-r)$
in the strong interacting regime $V_{0}\gg t$. We will see that the
results are insensitive to the particular form adopted for $V(r)$
as long it corresponds to a strong short-range potential.

\section{The dc conductivity of graphene and its bilayer \label{DCcond}}

As discussed in Sec.~\ref{introd}, there is growing evidence that
the limiting scattering mechanism of the electronic mobility in graphene
is due to strong short-range potentials, likely to have originated
from adsorbed hydrocarbons. These adsorbed atoms and/or molecules
act as resonant scatterers, giving rise to midgap states.\cite{robinson,stauberBZ,wehling,wehlingII}

This section is most important: it clarifies why the statement that
short-range scatterers in graphene give a dc conductivity independent
of the gate voltage is erroneous. As noted in Sec.~\ref{introd},
this misleading idea has its roots in the FBA, which fails blatantly
in this problem, as we demonstrate in what follows.

\subsection{Adsorbed atoms in graphene as strong short-range scattering centers\label{sec_resonant}}

The resonant scattering mechanism is easy to seize by considering
a simple model. Let us write the tight-binding Hamiltonian of the
$\pi$ electrons in graphene as (spin index omitted) \begin{equation}
H=-t\sum_{n,\bm{\delta}_{i}}\vert A,\bm{R}_{n}\rangle\langle\bm{R}_{n}+\bm{\delta}_{i},B\vert+\textrm{H.c.}\,,\label{eq_tbhamilt}\end{equation}
 where $\vert A,\bm{R}_{n}\rangle$ represents the Wannier state at
the unit cell $\bm{R}_{n}$; an equivalent definition holds for $\vert B,\bm{R}_{n}+\bm{\delta}_{i}\rangle$,
where $\bm{\delta}_{i}$ is one of three nearest-neighbor vectors
in the honeycomb lattice, as depicted in Fig.~\ref{fig_monolayer_lattice}.

We now consider that an impurity is binding covalently to a carbon
atom at site $\bm{R}_{n}=0$. This situation adds to the Hamiltonian
in Eq.~(\ref{eq_tbhamilt}) a term of the form \begin{equation}
H_{\textrm{rs}}=(V_{\textrm{ad}}\vert\textrm{ad}\rangle\langle A,0\vert+\textrm{h.c.})+\epsilon_{\textrm{ad}}\vert\textrm{ad}\rangle\langle\textrm{ad}\vert\,,\label{eq:rs_contribution_ham}\end{equation}
 where $V_{\textrm{ad}}$ is the hybridization between the adatom
(or a carbon atom of a hydrocarbon molecule) and a given carbon atom
of graphene, $\epsilon_{\textrm{ad}}$ is the relative (to graphene's
carbon atoms) on-site energy of the electron in the adatom, and $\vert\textrm{ad}\rangle$
is the ket representing the state of the electron in the adatom. Taking
the wave function to be of the form \begin{eqnarray}
\vert\psi\rangle & = & \sum_{n}[A(\bm{R}_{n})\vert A,\bm{R}_{n}\rangle+B(\bm{R}_{n}+\bm{\delta}_{2})\vert B,\bm{R}_{n}+\bm{\delta}_{2}\rangle]\nonumber \\
 & + & C_{\textrm{ad}}\vert\textrm{ad}\rangle\,,\end{eqnarray}
 the Schrödinger equation applied to the site $\bm{R}_{n}=0$ reads
\begin{eqnarray}
EA(0)-V_{\textrm{ad}}C_{\textrm{ad}} & = & -t[B(\bm{\delta}_{1})+B(\bm{\delta}_{2})+B(\bm{\delta}_{3})]\,,\\
\nonumber \\(E-\epsilon_{\textrm{ad}})C_{\textrm{ad}} & = & V_{\textrm{ad}}A(0)\,.\end{eqnarray}
 Solving for $C_{\textrm{ad}}$, we obtain \begin{equation}
-t[B(\bm{\delta}_{1})+B(\bm{\delta}_{2})+B(\bm{\delta}_{3})]=\left(E-\frac{V_{\textrm{ad}}^{2}}{E-\epsilon_{\textrm{ad}}}\right)A(0)\,.\label{eq:resonantTB}\end{equation}
 The resonant effect is included in the last term in Eq.~(\ref{eq:resonantTB}),
which represents an effective local potential of strength \begin{equation}
V_{\textrm{eff}}=V_{\textrm{ad}}^{2}/(E-\epsilon_{\textrm{ad}})\,.\label{eq_Veff}\end{equation}
 Quantum chemical calculations can determine the value of the parameters
$\epsilon_{\textrm{ad}}$ and $V_{\textrm{ad}}$.\cite{robinson,wehling,wehlingII}
Typical values are $V_{\textrm{ad}}\sim2t\sim5$ eV and $\epsilon_{\textrm{ad}}\sim-0.2$
eV,\cite{wehlingII} leading to $V_{\textrm{eff}}\sim100$ eV at half-filling
($E=0$), a rather strong on-site potential. On the basis of this
fact, it is natural to expect that adsorbates (i.e., resonant scatterers)
and vacancies lead to similar effects on the electronic structure
and transport properties. In monolayer graphene, vacancies are known
to significantly alter the density of states at energies close to
$\epsilon_{\textrm{ad}}$. In particular, vacancies induce a large
spectral transference from the Van Hove singularities to the neighborhood
of the Dirac point. As a consequence, the density of states displays
sharp peaks within the midgap region.\cite{vitorpaco,Castro2010}
This effect was first demonstrated in Ref.~\onlinecite{vitorpaco};
recently, it has been shown that indeed adsorbates do originate similar
behavior.\cite{wehlingII,yuan1}

\begin{figure}
\begin{centering}
\includegraphics[width=8cm]{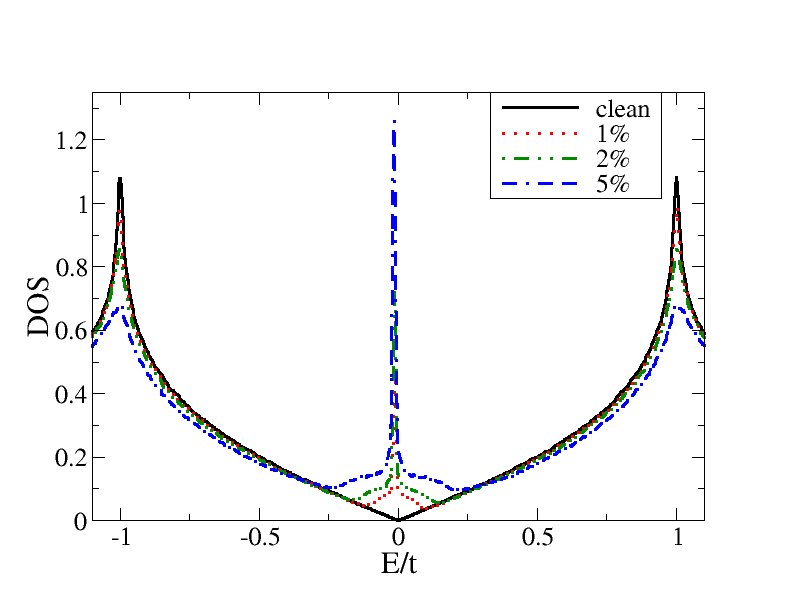} 
\par\end{centering}

\begin{centering}
\includegraphics[width=8cm]{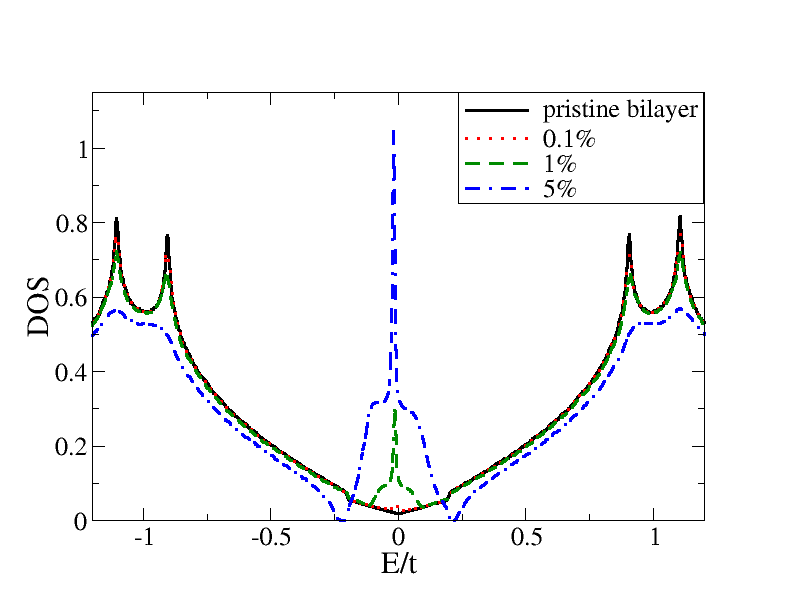} 
\par\end{centering}

\caption{\label{fig:KPM_DOS}(Color online) Effect of adatoms (resonant scatterers)
on the density of states (DOS) of monolayer graphene (top) and bilayer
graphene (bottom). Calculation of the DOS was carried out in honeycomb
lattices with $N=1000\times1000$ carbon sites for different concentrations
of adsorbed atoms (periodic boundary conditions and 10 realizations
of disorder were taken). The tight-binding parameters read $V_{\textrm{ad}}=2t$,
$\epsilon_{\textrm{ad}}=-0.0625t$, and $t_{\perp}=0.2t$. The DOS
discloses a dislocation of spectral weight toward the midgap region,
a phenomenon first reported for vacancies in monolayer graphene in
Ref.~\onlinecite{vitorpaco}.}

\end{figure}

\begin{figure}
\centering{}\includegraphics[width=8.5cm]{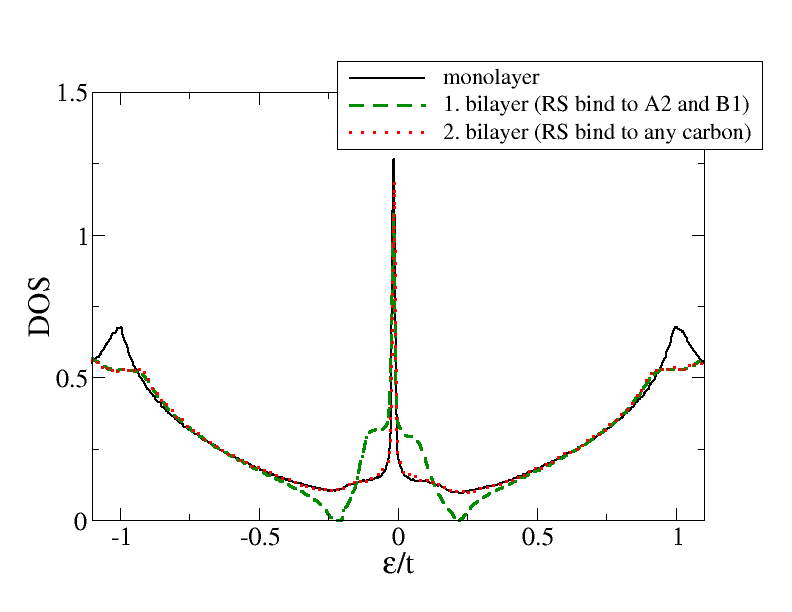} \caption{\label{fig:KPM_DOS_b}(Color online) Density of states (DOS) for bilayer
systems with $5\%$ resonant scatterers (RS) in the two scenarios
described in the text, namely, (1) adsorbates binding only to carbons
$A_{2}$ and $B_{1}$, and (2) adsorbates forming bonds with carbons
in any sublattice. The first situation opens a gap between the impurity
band and high-energy states. The DOS of monolayer graphene is shown
for comparison; tight-binding parameters are given in the caption
to Fig.~\ref{fig:KPM_DOS}.}

\end{figure}

Here we report similar results for bilayer graphene. To calculate
the density of states, we employ the KPM (see Ref.~\onlinecite{Weisse-Review}
for a review). For the sake of simplicity, we have considered equal
concentrations of adsorbates in both bottom and top layers. (The actual
applicability of this choice depends on the laboratory conditions
and specific experimental setup.) In what follows, we discuss the
situation where the adatoms bind only to carbons with coordination
number $z=3$ (i.e., those termed $A_{2}$ and $B_{1}$ in Fig.~\ref{fig_bilayer_lattice}).

The effect of resonant impurities in the electronic structure of monolayer
and bilayer graphene for different adsorbates concentrations, $n_{\textrm{ad}}$,
per carbon atom, is shown in Fig.~\ref{fig:KPM_DOS}. For illustration
purposes, we present the results for a high defect concentration,
$n_{\textrm{ad}}\sim1\%$, so that the modification of the graphene
electronic structure is visible to the eye in a wide energy window;
later we will see that the estimated values for defect concentration,
for typical experimental conditions, are actually far below these
values (Secs.~\ref{sec_boltzmann_GSL} and \ref{sec_boltzmann_GBL}).

In both graphene systems, the adatoms lead to well-defined peaks close
to zero energy, the so-called midgap region. As mentioned above, such
enhancement of the density of states is accompanied by a decrease
in spectral weight near the Van Hove singularities, a situation reminiscent
of vacancy-induced disorder.\cite{vitorpaco,Castro2010} The effective
potential {[}Eq.~(\ref{eq_Veff}){]}, despite being very strong,
is bounded, explaining the slight electron-hole asymmetry near the
Dirac point. The resonant peaks are centered at negative energies
because $\epsilon_{\textrm{ad}}<0$. Increasing the impurity concentration
brings more spectral weight toward the midgap region. In bilayer graphene,
though, a curious phenomenon takes place: when the impurity concentration
is large enough, a gap opens separating midgap states, forming the
impurity band, from high energy states (see Fig.~\ref{fig:KPM_DOS},
bottom). Similar findings were reported in recent \emph{ab initio}
calculations considering asymmetric doping of graphene.\cite{Menezes}

Figure~\ref{fig:KPM_DOS_b} shows how the electronic structure changes
when the restriction on the allowable carbon-impurity bonds is relaxed.
When adsorbates bind to carbons in any sublattice in the bilayer,
the density of states is almost indistinguishable from that of monolayer
graphene (with the same impurity concentration). The latter is accurate
for a large energy window around the Dirac point ($|\epsilon|\lesssim0.5t$);
for higher electronic energies, the density of states becomes insensitive
to the type of impurity-carbon bonds present in the bilayer samples.
Roughly speaking, forming chemical bonds to every type of carbon decouples
the layers, and hence dc-transport properties will be similar to those
of a single layer of graphene (Sec.~\ref{sub:KPM}).

In light of the present results and previous reports for vacancies\cite{vitorpaco,Castro2010}
and resonant impurities in monolayer graphene,\cite{wehlingII,yuan1}
we are led to conclude that the formation of an impurity band in the
midgap region is universal in graphene systems with typical adsorbed
species. In Sec.~\ref{sub:KPM}, it will be shown that such an impurity
band has a strong impact on the transport properties of undoped graphene.

Away from neutrality, the calculation of transport properties for
the effective local potential model can be performed using the $T$
matrix approach.\cite{tsai1,tsai2,robinson} Its derivation for resonant
scatterers is elementary. It is well known that the $T$ matrix for
a local potential of intensity $v_{0}$ reads \cite{bena,nmrPRB06}
\begin{equation}
T(E)=v_{0}[1-v_{0}\bar{G}_{R}(E)]^{-1}\,,\end{equation}
 and \begin{equation}
\bar{G}_{R}(E)=ED^{-2}\ln(E^{2}/D^{2})-i\pi\vert E\vert/D^{2}\,,\end{equation}
 with $D\simeq3t$. Then, using Eq.~(\ref{eq_Veff}), the $T-$matrix
due to an adatom must be of the form \begin{equation}
T(E)=\frac{V_{\textrm{eff}}}{1-V_{\textrm{eff}}\bar{G}_{R}(E)}=\frac{V_{\textrm{ad}}^{2}}{E-\epsilon_{\textrm{ad}}-V_{\textrm{ad}}^{2}\bar{G}_{R}(E)}\,.\label{eq:rstmatrix}\end{equation}
 Since we are considering that $V_{\textrm{ad}}\gg(\epsilon_{\textrm{ad}},\vert E\vert)$,
we can approximate the $T$ matrix, Eq.~(\ref{eq:rstmatrix}), by
\begin{equation}
T(E)\approx-\frac{1}{\bar{G}_{R}(E)}\,,\label{eq:rstmatrix_vacancy}\end{equation}
 which is nothing but the $T$ matrix for vacancies.\cite{nmrPRB06}

The transport relaxation time $\tau(k_{F})$ (at the Fermi surface)
can be calculated using Fermi's golden rule, \begin{equation}
\hbar/\tau(k_{F})=\pi n_{i}^{c}\vert T(\epsilon_{F})\vert^{2}\rho(\epsilon_{F})\,,\label{eq:relaxation_time_golden_rule}\end{equation}
 where $n_{i}^{c}$ is the concentration of impurities per unit cell,
and $k_{F}$ and $\epsilon_{F}$ are the Fermi momentum and energy,
respectively. From the knowledge of $\tau(k_{F})$, the conductivity
of graphene follows from Boltzmann's transport equation (see the following
section).\cite{ziman}

\subsection{The Boltzmann approach to dc conductivity using partial-wave expansion\label{sub:The-Boltzmann-approach}}

The above analysis made transparent that the effect of resonant scatterers
is equivalent to that of a strong on-site potential (as long as the
$T$-matrix formalism is applicable). We can then use the formalism
of Sec.~\ref{partialwave} to compute the exact phase shifts in the
presence of such a strong potential, from which $\tau(k_{F})$ can
be obtained. This type of calculations is equivalent, and alternative,
to calculations based on the T-matrix approach in the lattice, with
the appropriate choice of the effective size of the impurity.

A relation between $\tau(k_{F})$ and $\sigma(\theta)$ is provided
by \cite{ziman} \begin{equation}
1/\tau(k_{F})=n_{i}\left(\mathbf{v}_{k_{F}}\cdot\mathbf{e}_{r}\right)\sigma_{\textrm{T}}\,,\label{eq:relaxation
  time}\end{equation}
 where $n_{i}$ is the concentration of impurities per unit area,
$\mathbf{v}_{k_{F}}$ is the velocity of the electrons at the Fermi
surface, $\mathbf{e}_{r}$ is the radial versor in cylindric coordinates,
and $\sigma_{T}$ is the total transport cross section,\cite{ziman}
\begin{eqnarray}
\sigma_{\textrm{T}} & = & \int_{0}^{2\pi}d\theta\,(1-\cos\theta)\sigma(\theta)\,\label{eq:sigma_T}\\
 & = & \frac{2}{k}\sum_{m=-\infty}^{\infty}\sin^{2}(\delta_{m}-\delta_{m+1})\equiv\frac{2}{k}\Lambda(k)\,.\label{eq_crosstransp}\end{eqnarray}
 The conductivity of a given material follows from Boltzmann's transport
equation. The electric current has the general form \begin{equation}
\bm{j}=\frac{g_{s}g_{v}e^{2}}{(2\pi)^{2}}\int d\mathbf{k}\,\tau(k)\frac{\partial n_{F}(k)}{\partial\varepsilon_{k}}(\mathbf{v}_{k}\cdot\mathbf{E})\mathbf{v}_{k}\,,\label{eq_current}\end{equation}
 where $n_{F}$ is the Fermi distribution function, $\varepsilon_{k}$
is the dispersion of the electron, $\mathbf{v}_{k}$ is the velocity
of the particle with momentum $k$, $\mathbf{E}$ is the external
electric field, and $g_{s}$ and $g_{v}$ are the spin and valley
degeneracies, respectively. The electron velocity at the Fermi surface
reads \begin{equation}
\mathbf{v}_{k_{F}}=v_{F}\mathbf{e}_{r}\,,\label{eq_VFgraphene}\end{equation}
 whereas in the bilayer it has the form \begin{equation}
\mathbf{v}_{k_{F}}=\frac{2v_{F}^{2}}{t_{\perp}}\hbar k_{F}\mathbf{e}_{r}\,,\label{eq_VFBLgraphene}\end{equation}
 which depends on the position of the Fermi energy; the quantity $M^{-1}=2v_{F}^{2}/t_{\perp}$
plays the role of the electron's band mass. The dc-conductivity, $\sigma_{\textrm{dc}}$,
can be obtained from Ohm's law, $j_{x}=\sigma_{\textrm{dc}}E_{x}$.
Combining Eqs.~(\ref{eq_crosstransp}), (\ref{eq_current}), (\ref{eq_VFgraphene}),
and (\ref{eq_VFBLgraphene}), the dc-conductivity for both monolayer
and bilayer graphene has one and the same form, namely, \begin{equation}
\sigma_{\textrm{dc}}=\frac{4e^{2}}{h}\frac{k_{F}^{2}}{4n_{i}\Lambda(k_{F})}\,,\label{eq_sigmaDC}\end{equation}
 where the zero-temperature limit has been taken. The importance of
Eq.~(\ref{eq_sigmaDC}) could not be more emphasized, since it shows
that the final dependence of the conductivity on $k_{F}$, and therefore
on the electronic density, is controlled by the behavior of $\Lambda(k_{F})$,
which depends only on the phase shifts $\delta_{m}$; these, in turn,
depend on the nature of the scattering potential. Therefore, the exact
calculation of the phase shifts emerges as the central theoretical
problem regarding the description of the variation of $\sigma_{\textrm{dc}}$
with the gate voltage for monolayer and bilayer graphene.

\subsection{Graphene \label{sec_boltzmann_GSL}}

For monolayer and bilayer, the electronic doping is controlled by
a backgate voltage $V_{g}$. The value of the Fermi momentum depends
on the density of electrons, and, therefore, also on $V_{g}$. If
the dielectric between graphene (or its bilayer) and the backgate
is made of silicon oxide and has a width of about 300 nm, then we
have \begin{equation}
k_{F}^{2}=\pi\alpha V_{g}\,,\end{equation}
 with $\alpha\simeq7.2\times10^{10}$ V$^{-1}\cdot$cm$^{-2}$; numerically
we have $k_{F}=4.7\times10^{-3}\times\sqrt{V_{g}}$\, \AA{}$^{-1}$.

As we have discussed in Sec.~\ref{sec_resonant}, an adsorbed atom
or molecule (of specific types) can be described as an effective strong
short-range potential. As a consequence, we model the effect of an
adsorbed (resonant) chemical specie at the surface of graphene by
a potential of the form

\begin{equation}
V(r)=V_{0}\theta(R-r)\,,\label{eq_potential}\end{equation}
 where $R$ has to be of the order of $\sim$1\AA{}\, and $V_{0}\gg t$.
As a limiting behavior, we consider that $V_{0}$ is made arbitrarily
large. In the Appendix we discuss the case where the potential is
represented by a Dirac delta-function. The latter problem can be solved
nonperturbatively, and an effective length scale $R_{{\rm \textrm{eff}}}$
emerges in the problem due to an energy cutoff associated with the
bandwidth. This effective length scale ($R_{{\rm \textrm{eff}}}$)
is identified with the range $R$ of the potential given above. Both
problems lead to the same results for the conductivity of graphene
(see later).

In the limit $V_{0}\rightarrow\infty$, the potential defines an impenetrable
barrier to the electronic probability flux. For electrons described
either by the Schrödinger equation or by the Hamiltonian in Eq.~(\ref{eq_BL2comp}),
the condition of zero flux for $r<R$ is achieved by imposing that
$\Psi(r=R)=0$ {[}$\Psi(r)$ represents either a scalar or a spinor{]}.
For electrons described by the massless Dirac equation, the latter
implies that the wave function has to vanish everywhere and, therefore,
cannot be used. In contrast, from Eq.~(\ref{eq_JrDirac}) it is clear
that the radial flux at $r=R$ can be made 0 if one of the components
of the spinor is 0 at $r=R$. \cite{adame} In conclusion, the correct
boundary condition enforcing zero flux at $r=R$ for electrons in
monolayer graphene is given by \begin{equation}
\Psi_{i}(r=R)=0\,,\label{eq_zerofluxDirac}\end{equation}
 where $\Psi_{i}$, with $i=1,2$, is one of the components of the
spinor. Given the presence of two Dirac cones in graphene, it is immaterial
which component we choose to obey the condition of Eq.~\ref{eq_zerofluxDirac}),
as long as we consider the contributions to the two Dirac cones in
the Brillouin zone of the honeycomb lattice.

\begin{figure}[ht]
 \centering{} \includegraphics[width=8cm]{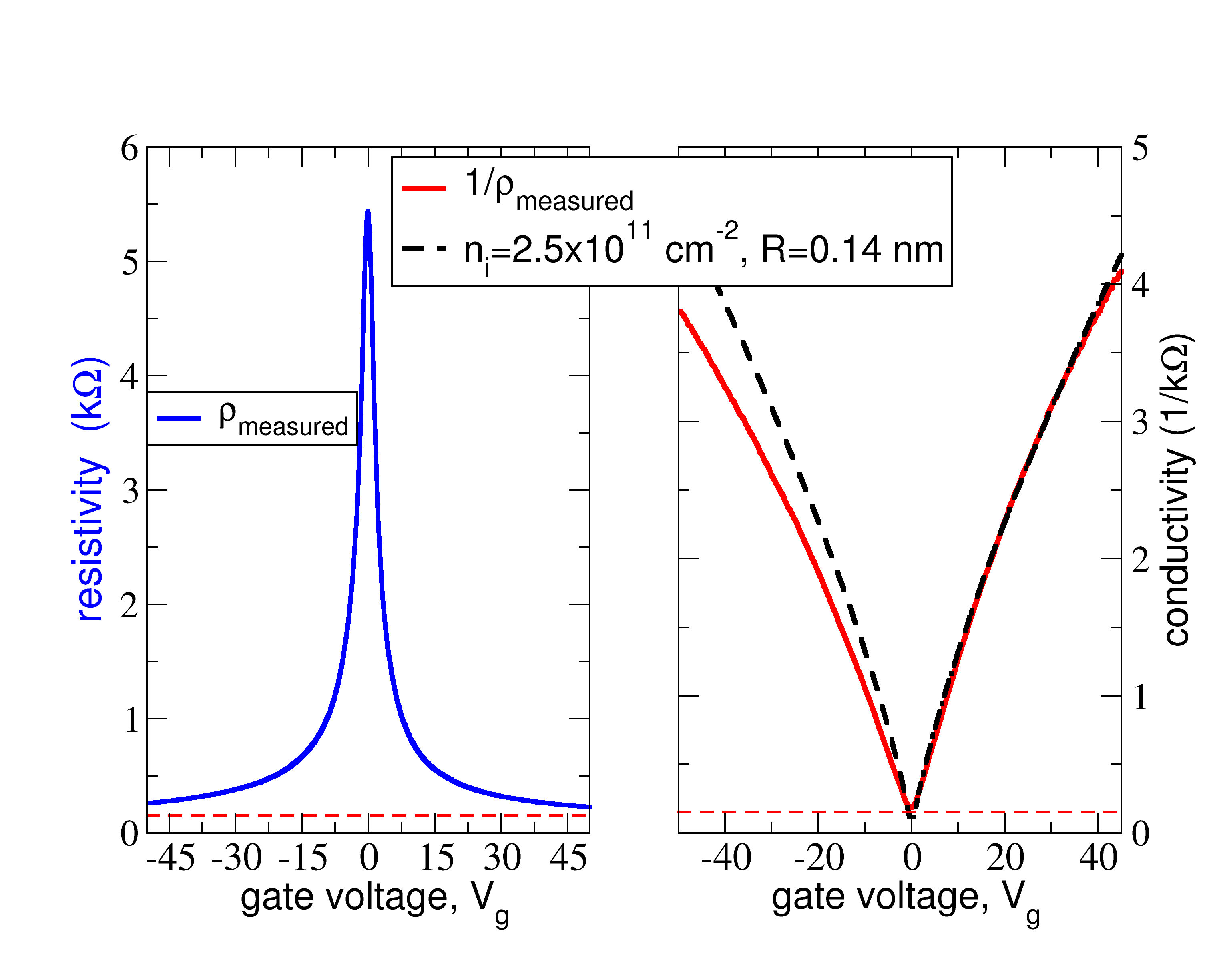} \caption{\label{fig_graphene_data}(Color online) Experimental data on graphene's
conductivity. Left: raw data on a measurement of the resistivity,
$\rho_{\textrm{measured}}$, of an exfoliated graphene sheet. Right:
Fit of the conductivity, $\sigma_{\textrm{sub}}=1/\rho_{\textrm{measured}}$,
using Eq.~(\ref{eq_sigma_graphene}). The value of $R$ was taken
to be of the order of $a_{0}$ and the fit provided an areal density
of impurities of $n_{i}\approx2.5\times10^{11}$ cm$^{-2}$ (or a
concentration $n_{\textrm{ad}}\approx5\times10^{-5}$ per carbon atom).
In both panels, the horizontal dashed line (red) stands for twice
the quantum of conductance, that is, $2e^{2}/h$. (Data from S. V.
Morozov \textit{et al.},\cite{morozov} courtesy of A. K. Geim.)}

\end{figure}

To satisfy the boundary condition in Eq.~(\ref{eq_zerofluxDirac}),
we write the wave function describing the electrons being scattered
by the barrier as \begin{equation}
\Psi_{m}(r,\theta)=A_{1}^{m}\left[\begin{array}{rc}
J_{m}(kr)\\
e^{i\theta}J_{m+1}(kr)\end{array}\right]+A_{2}^{m}\left[\begin{array}{rc}
Y_{m}(kr)\\
e^{i\theta}Y_{m+1}(kr)\end{array}\right]\,.\label{eq_waveDirac_Well}\end{equation}
 Thus, the boundary condition in Eq.~(\ref{eq_zerofluxDirac}) implies
that \begin{equation}
\frac{A_{2}^{m}}{A_{1}^{m}}=-\frac{J_{m}(kR)}{Y_{m}(kR)}\,.\label{eq:a2m_a1m}\end{equation}
 Since for large $r$, the wave function in Eq.~(\ref{eq_waveDirac_Well})
must have the general form shown in Eq.~(\ref{eq_psi_dirac_angular}),
it follows that the ratio $A_{2}^{m}/A_{1}^{m}$ has to be interpreted
as \begin{equation}
\frac{A_{2}^{m}}{A_{1}^{m}}=-\tan\delta_{m}\,,\label{eq_ratioDirac}\end{equation}
 which defines the phase shift $\delta_{m}$. {[}A comment about the
latter result is in order: In graphene, radially symmetric potentials
originate phase shifts obeying $\delta_{m}=\delta_{-m-1}$. This can
be seen by noting that replacing $m$ by $-m-1$ in Eq.~(\ref{eq:solution_dirac_slg})
produces another eigenstate of the Dirac Hamiltonian. Equations (\ref{eq:a2m_a1m})
and (\ref{eq_ratioDirac}) show that impenetrable barriers force a
different symmetry: $\delta_{m}=\delta_{-m}$.{]} For backgate voltage
values in the range $V_{g}\lesssim100$ V, and considering $R\sim1$\AA{}\,,
we have $Rk<1$ (known as the low-energy scattering regime). In this
regime, the scattering is dominated by the $s$ wave phase shift;
that is, the dominant contribution to $\Lambda(k)$ comes from \begin{equation}
\tan\delta_{0}=\frac{J_{0}(kR)}{Y_{0}(kR)}\approx\frac{\pi}{2}\ln^{-1}(kR)\,,\label{eq_delta0_pha_shif}\end{equation}
 where Eqs.~(\ref{eq_Jm_smallx}) and (\ref{eq_Y0_smallx}) have
been used. It follows from Eqs.~(\ref{eq_crosstransp}) and (\ref{eq_delta0_pha_shif})
that the conductivity of graphene obtained from Eq.~(\ref{eq_sigmaDC})
has the final form \cite{nmrPRB06,OstrovskySCBA,stauberBZ,basko}
\begin{equation}
\sigma_{\textrm{dc}}=\frac{4e^{2}}{h}\frac{k_{F}^{2}}{2\pi^{2}n_{i}}\ln^{2}(k_{F}R)\,.\label{eq_sigma_graphene}\end{equation}
 Given that the value of $R$ is constrained to be of the order of
$1$ \AA{}\,, $n_{i}$ is the only fitting parameter. Equation (\ref{eq_sigma_graphene})
was used to fit the conductivity data \cite{morozov} of an exfoliated
graphene sheet, as shown in Fig.~\ref{fig_graphene_data}. Because
we took the limit $V_{0}\rightarrow\infty$, the computed conductivity
does not break electron-hole symmetry. The electron-hole asymmetry
shown by the experimental data in Fig.~\ref{fig_graphene_data} can
be attributed to the presence of charge scatterers and/or to the role
of the contacts.\cite{Goldhaber} If we increase the value of $R$
somewhat, the concentration of impurities needed to fit the data decreases.
In Fig.~\ref{fig_graphene_data} we have chosen to fit the conductivity
for a positive gate voltage; it is manifest that Eq.~(\ref{eq_sigma_graphene})
fits the data accurately {[}dashed (black) curve{]}. If we had decided
to fit the data for negative values of $V_{g}$, the obtained concentration
of impurities, $n_{i}$, would have been slightly different. The concentration
of scatterers is rather small (see caption to Fig.~\ref{fig_graphene_data})
and agrees with the concentration of atomic scale defects estimated
via Raman measurements. \cite{Dpeak} This testifies to the strong
effect of a few resonant scatterers dilluted in the surface of graphene
(similar to atomic vacancies), as discussed in Sec.~\ref{sec_resonant}.

The result given by Eq.~(\ref{eq_sigma_graphene}) for the conductivity
of monolayer graphene can also be obtained from a model where vacancies
act as scattering centers.\cite{nmrPRB06} In view of the arguments
given in Sec.~\ref{sec_resonant}, this result comes as no surprise,
since the effective local potential created by adsorbed hydrocarbons
is much larger than the hopping integral $t$. Numerical simulations
of the dc conductivity based on Kubo's formula in the presence of
local potentials found a sublinear behavior for a graphene monolayer,\cite{nomura}
in qualitative agreement with Eq.~(\ref{eq_sigma_graphene}).

Let us now extend the previous analysis to the case of a graphene
bilayer.

\subsection{Graphene bilayer \label{sec_boltzmann_GBL}}

Assuming that the dominant source of scattering in graphene is due
to strong short-range potentials, then the same must be true for bilayer
graphene. As a consequence, a consistent description of electronic
scattering in both monolayer and bilayer graphene must use the same
scattering potential to explain the measured conductivity in both
systems. In the spirit of this work, this means that the scattering
potential in Eq.~(\ref{eq_potential}) must also be used to compute
the conductivity of graphene bilayer.

\begin{figure}[ht]
 \centering{} \includegraphics[width=8cm]{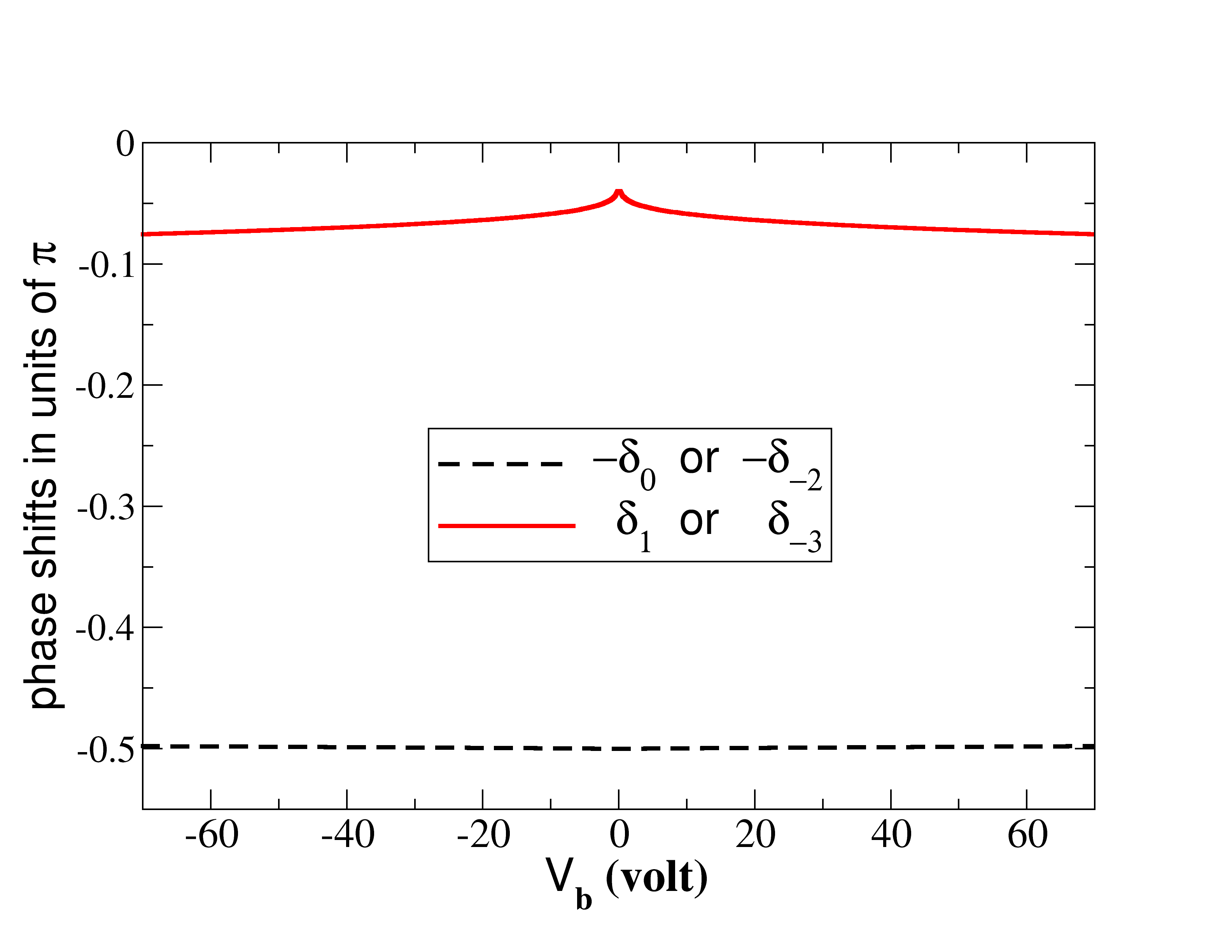} \caption{\label{fig_phase_shift_01} (Color online) Dependence of the phase
shifts $\delta_{0(-2)}$ (solid line) and $\delta_{1(-3)}$ (dashed
line) on $V_{g}$, for bilayer graphene with $R=a_{0}$. The differences
between the exact expressions in Eq.~(\ref{eq_tangdel0A1A2BL}) and
the asymptotic values in Eqs.~(\ref{eq:tan_d0}) and (\ref{eq:tan_d1})
are not visible to the eye. Other phase shifts are approximately 0
within the same range of $V_{g}$.}

\end{figure}

As in the case of Eq. (\ref{eq_waveDirac_Well}), we seek a wave function
in the form of a superposition of Bessel functions of different kinds,
which in the present case assumes the form \begin{eqnarray}
\Psi_{m}(r,\theta) & = & A_{1}^{m}\left[\begin{array}{c}
J_{m}(kr)\\
-e^{2i\theta}J_{m+2}(kr)\end{array}\right]\nonumber \\
 & + & A_{2}^{m}\left[\begin{array}{c}
Y_{m}(kr)\\
-e^{2i\theta}Y_{m+2}(kr)\end{array}\right]\nonumber \\
 & + & A_{3}^{m}\left[\begin{array}{c}
K_{m}(kr)\\
-e^{2i\theta}K_{m+2}(kr)\end{array}\right].\label{eq_waveBL_Well}\end{eqnarray}
 The introduction of the modified Bessel function $K_{m}(kr)$ in
Eq.~(\ref{eq:tight-binding-Hamilt-BLG}) is necessary to satisfy
the boundary condition $\Psi(r=R)=0$. We recall that Hamiltonian
in Eq.~(\ref{eq_BL2comp}) supports evanescent waves at the boundary
$r=R$, as discussed in Sec.~\ref{subsecBL}. Furthermore, for large
$r$, $K_{m}(kr)$ decays exponentially, as we can see from Eq.~(\ref{eq_Km_largex}).
Therefore, at large distances, the behavior of the wave function in
Eq.~(\ref{eq:tight-binding-Hamilt-BLG}) depends only on the form
of $J_{m}(kr)$ and $Y_{m}(kr)$, as given by Eqs.~(\ref{eq_BesselJ_xlarge})
and (\ref{eq_BesselY_xlarge}). As a consequence, the phase shift
$\delta_{m}$ is determined by the ratio $A_{2}^{m}/A_{1}^{m}$; that
is, we must have \begin{equation}
\frac{A_{2}^{m}}{A_{1}^{m}}=-\tan\delta_{m}\,,\label{eq_tangdel0A1A2BL}\end{equation}
 as in the case of electrons in monolayer graphene {[}see Eq.~(\ref{eq_ratioDirac}){]}.
Imposing the boundary condition $\Psi(r=R)=0$ on the wave function
(\ref{eq:tight-binding-Hamilt-BLG}), we obtain \begin{eqnarray}
0 & = & A_{1}^{m}J_{m}(kR)+A_{2}^{m}Y_{m}(kR)+A_{3}^{m}K_{m}(kR)\,,\\
0 & = & A_{1}^{m}J_{m+2}(kR)+A_{2}^{m}Y_{m+2}(kR)+A_{3}^{m}K_{m+2}(kR)\,,\nonumber \end{eqnarray}
 from which follows \begin{equation}
\frac{A_{2}^{m}}{A_{1}^{m}}=\frac{J_{m}(kR)K_{m+2}(kR)-J_{m+2}(kR)K_{m}(kR)}{K_{m}(kR)Y_{m+2}(kR)-K_{m+2}(kR)Y_{m}(kR)}\,.\label{eq_ratioA1A2BL}\end{equation}
Combining Eqs.~(\ref{eq_tangdel0A1A2BL}) and (\ref{eq_ratioA1A2BL}),
the equation for the phase shift $\delta_{m}$ follows at once. Contrary
to the case of monolayer graphene, the cross section is no longer
dominated by $\delta_{0}$ alone. The asymptotic expansions for $\delta_{0}$
and $\delta_{1}$ are ($k_{F}R<1$) \begin{equation}
\tan\delta_{0}=-\frac{\pi}{2(k_{F}R)^{2}}[\ln(k_{F}R/2)+\gamma_{E}-1/2]^{-1},\label{eq:tan_d0}\end{equation}
 and \begin{equation}
\tan\delta_{1}=\frac{\pi}{4}[\ln(k_{F}R/2)+\gamma_{E}-1/4]^{-1},\label{eq:tan_d1}\end{equation}
 where $\gamma_{E}=0.577\ldots$ is Euler's constant. In addition,
we have two more nonzero phase shifts: \begin{equation}
\delta_{-2}=\delta_{0},\;\text{and}\;\delta_{-3}=\delta_{1}.\label{eq:delta_other}\end{equation}
 These expressions are exact and reflect a symmetry of the eigenstates
of Eq.~(\ref{eq_BL2comp}) when radially symmetric scalar potentials
are considered, namely, $\delta_{m}=\delta_{-m-2}$.

\begin{figure}[ht]
 \centering{} \includegraphics[width=8cm]{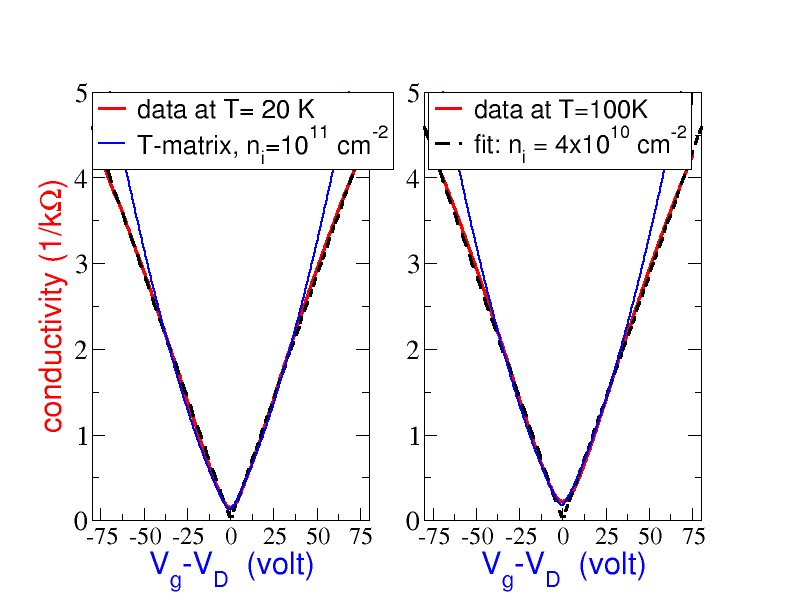} \caption{\label{fig_graphene_data_BL}(Color online) Fit of the conductivity
data of bilayer graphene {[}solid (red) curve{]} using Eqs.~(\ref{eq_sigma_BL})
and (\ref{eq_nilsson_sigma}). The fit has only a single parameter,
the concentration of impurities. The obtained value is $n_{i}\approx4\times10^{10}$
cm$^{-2}$ ( concentration $n_{\textrm{ad}}\approx1\times10^{-5}$
per carbon atom), for Eq.~(\ref{eq_sigma_BL}), and $n_{i}\approx1\times10^{11}$
cm$^{-2}$ (concentration $n_{\textrm{ad}}\approx0.25\times10^{-5}$
per carbon atom) for the $T$ matrix approach, using a model of pure
vacancies, Eq.~(\ref{eq_nilsson_sigma}). Left: Data taken at a temperature
of 20 K. Right: Conductivity of the same sample at the higher temperature
of 100 K. The position of the Dirac point, $V_{D}$, was shifted to
0 in this figure. (Data from S. V. Morison \textit{et al.,}\cite{morozov}
courtesy of A. K. Geim.)}

\end{figure}

The dependence of $\delta_{0}$ and $\delta_{1}$ on $V_{g}$ is given
in Fig. \ref{fig_phase_shift_01}. From Eqs.~(\ref{eq:tan_d0})--(\ref{eq:delta_other}),
it follows that $\Lambda(k_{F})\simeq4$. The dc conductivity of bilayer
graphene is, therefore, given by \begin{equation}
\sigma_{\textrm{dc}}=\frac{4e^{2}}{h}\frac{k_{F}^{2}}{16n_{i}}\,.\label{eq_sigma_BL}\end{equation}
 Curiously, the symmetry of the scattering amplitudes combine to make
$\Lambda(k_{F})$ independent of $k_{F}$ (with an accuracy better
than $1$\% in the relevant range of $k_{F}$ and $R$), making the
conductivity proportional to the gate voltage. This result, together
with the constant density of states (valid when $\vert E\vert\ll t_{\perp}$),
is at the heart of the exact linear dependence of the conductivity
on the gate-voltage. We have used Eq.~(\ref{eq_sigma_BL}) to fit
the conductivity data of an exfoliated bilayer graphene sample, as
shown in Fig.~\ref{fig_graphene_data_BL}. The fit provides a concentration
of impurities of the order of $n_{i}\approx4\times10^{10}$ cm$^{-2}$
(i.e., a concentration of adatoms per carbon atom of about $n_{\textrm{ad}}\approx1\times10^{-5}$).
Since in bilayer graphene only two of the four surfaces are exposed
to the environment, the $n_{i}$ value found above, being slightly
smaller than that found for monolayer graphene, can be interpreted
as a manifestation of this fact.

Within the $T$ matrix approach, the dc conductivity of bilayer graphene
has been computed in the past.\cite{nilsson1,nilsson2} The impurity
concentrations used in those works were far too large to reveal the
linear behavior in $V_{g}$ given by Eq.~(\ref{eq_sigma_BL}). We
have already shown that the effect of resonant scatterers can be captured
by a model of pure vacancies, using both the $T$ matrix and the partial-wave
approaches. We now revisit the $T$ matrix calculation in bilayer
graphene \cite{nilsson1,nilsson2,Koshino} and show that, as in the
case of the monolayer, a model of pure vacancies in the bilayer also
captures the physics of resonant scatterers.

\subsection{$T$ matrix approach for bilayer graphene \label{sub:tmatrix}}

In Refs.~\onlinecite{nilsson1} and \onlinecite{nilsson2}, the
calculation of the dc conductivity took into account the full band
structure of the graphene bilayer. That calculation could distinguish
the four carbon atoms in the unit cell. In this section, we assume
that vacancies are located at the two carbons that are not coupled
by $t_{\perp}$.

In the notation in Refs.~\onlinecite{nilsson1},~\onlinecite{nilsson2},
the zero-temperature dc conductivity obtained from Kubo's formula
is given by \begin{eqnarray}
\sigma_{\textrm{dc}} & = & \frac{8e^{2}}{\pi h}\int_{0}^{\Lambda^{2}}d(k^{2})\left\{ \textrm{Im}[g_{AA}^{\textrm{D}}(E_{F},k)]\textrm{Im}[g_{BB}^{\textrm{D}}(E_{F}+\delta,k)]\right.\nonumber \\
 & + & \left.\textrm{Im}[g_{AB}^{\textrm{ND}}(E_{F},k)]\textrm{Im}[g_{AB}^{\textrm{ND}}(E_{F}+\delta,k)]\right\} \end{eqnarray}
 in the limit $\delta\rightarrow0$; see Ref.~\onlinecite{nilsson2}
for the definitions of the Green's functions $g(E,k)$. The $k^{2}$-integral
can be performed exactly, as explained in Appendix C in Ref.~\onlinecite{nilsson2}.
The resulting complicated formula can be approximated by going through
the following steps: (i) neglect the real part of the self-energies,
(ii) expand the result in powers of the imaginary part of the self-energies
$\Gamma_{a}(\e)\equiv-\text{Im}[\Sigma_{a}(\e)]$, and (iii) assume
that the energies involved fulfill $|\mu|,t_{\perp}\pm|\mu|\gg\Gamma_{A}(\e),\,\Gamma_{B}(\e)$.
The leading term in this expansion yields the approximate formula
\begin{equation}
\sigma_{\textrm{dc}}\approx\frac{2e^{2}}{h}\frac{E_{F}(E_{F}+t_{\perp})}{t_{\perp}\Gamma_{B}(E_{F})+E_{F}[\Gamma_{A}(E_{F})+\Gamma_{B}(E_{F})]}.\label{eq:approx_bl_conductivity1}\end{equation}
 This expression is a good approximation for low impurity concentrations
and away from the neutrality point, where the condition in step iii
breaks down. This result may be further simplified using the relation
between the Fermi energy and the density (assuming $n,E_{F}>0$) coming
from the dispersion relation $E_{F}=\sqrt{(t_{\perp}/2)^{2}+\pi(\hbar v_{F})^{2}n}-t_{\perp}/2$,
resulting in \begin{equation}
\sigma_{\textrm{dc}}=\frac{2e^{2}}{h}\frac{\pi(\hbar v_{F})^{2}n}{t_{\perp}\Gamma_{B}(E_{F})+E_{F}[\Gamma_{A}(E_{F})+\Gamma_{B}(E_{F})]}\,,\label{eq_nilsson_sigma}\end{equation}
 where $n$ is the electronic density. To the extent that the denominator
is independent of $E_{F}$, the conductivity is linear in the density
of carriers, $n$, in agreement with the description based on the
phase shifts. For low impurity densities, as is the case in exfoliated
samples, the difference between the conductivity obtained from the
coherent potential approximation and the $T$ matrix is very small
except in a tiny region near the neutrality point. Using Eqs.~(\ref{eq_sigma_BL})
and (\ref{eq_nilsson_sigma}), the data in Fig.~(\ref{fig_graphene_data_BL})
can be reasonably fit considering a density of vacancies of $n_{i}\simeq10^{11}$
cm$^{-2}$.

\subsection{Exact amplitudes versus first Born approximation\label{sub:delta}}

The use of the FBA within the semiclassical Boltzmann approach is
a common practice in condensed matter. In the present context, the
FBA has been employed to investigate the interplay between short-range
and long-range scattering.\cite{Sarma2010,Adam2008} Its use, however,
requires the weak scattering condition to be verified. We have seen
in Sec.~\ref{sec_resonant} that adsorbed atoms in graphene give
rise to strong local potentials $V_{0}\gg t$ , rendering inappropriate
the use of the FBA for a description of scattering due to realistic
short-range potentials.

The form of the graphene conductivity {[}see Eq.~(\ref{eq_sigma_graphene}){]}
is not peculiar when hard-wall boundary conditions are present; potentials
characterized by delta functions in real space yield equivalent results
if exact scattering amplitudes are considered instead of the FBA (see
Appendix). Moreover, beyond Boltzmann's kinetic theory, tight-binding
calculations for graphene sheets with $\sim0.02$ $\mu$m$^{2}$ show
quantitative agreement with Eq.~(\ref{eq_sigma_graphene}), while
at the same time displaying qualitative disagreement with the FBA.\cite{Klos2010}

To demonstrate that $\delta$ potentials also mimic the effect of
strong range potentials, we calculate the exact scattering cross sections
using the Lippmann-Schwinger equation, an approach well suited to
$\delta$ potentials. (To the best of our knowledge, the case of bilayer
graphene has not been considered before.) The calculations are shown
in the Appendix and important limiting cases are summarized in Table~\ref{tab:comparasion exact and born}.

\begin{table}
\begin{centering}
\begin{tabular}{lcc}
\hline 
dc-conductivity  & monolayer  & bilayer\tabularnewline
\hline 
$\delta$  &  & \tabularnewline
$\quad$FBA & Const.  & $\sim k_{F}^{2}$\tabularnewline
$\quad$nonperturbative; large$V_{0}$  & $\sim\left[k_{F}\ln\left(k_{F}R\right)\right]^{2}$  & $\sim k_{F}^{2}$\tabularnewline
Hard-disk radius $R$  & $\sim\left[k_{F}\ln\left(k_{F}R\right)\right]^{2}$  & $\sim k_{F}^{2}$\tabularnewline
\hline
\end{tabular}
\par\end{centering}

\caption{\label{tab:comparasion exact and born} The conductivity due to a
delta $\delta$ potential: the FBA and the nonperturbative result
in the relevant regime $V_{0}\gg$ all scales. For comparison, the
hard-disk result is listed. Although for the bilayer both the FBA
and the exact calculation give a conductivity proportional to $k_{F}^{2}$,
we should note that in the former case the conductivity is proportional
to the strength of the potential, and therefore the FBA cannot be
trusted in the regime of strong potentials, and the agreement of the
two approaches is fortuitous.}

\end{table}

For monolayer graphene the conductivity due to a $\delta$ potential
with strength $V_{0}$ reads \begin{equation}
\sigma_{\textrm{dc}}=\frac{4e^{2}}{h}\frac{2}{n_{i}}\left[(k_{F}/2\pi)\ln\left(k_{F}R\right)-\frac{\hbar v_{F}}{V_{0}}\right]^{2},\label{eq:sigmaDC_delta_SLG}\end{equation}
 where $R$ is a length scale introduced to regularize the Green's
function. The FBA is recovered from Eq.~(\ref{eq:sigmaDC_delta_SLG})
by considering $V_{0}$ smaller than relevant scales, yielding a conductivity
that does not depend on the carrier density/gate voltage. In contrast,
the strong scattering limit $V_{0}\gg\vert E\vert$ gives the same
dependence found for the hard disk model {[}Eq.~(\ref{eq_sigma_graphene}){]}
upon the identification of $R$ with the potential range.

The situation is quite different in bilayer graphene being described
by a low-energy theory of massive electrons: both weak and strong
scattering regimes yield a conductivity proportional to $k_{F}^{2}$
in the entire carrier density range; the exact result reads \begin{equation}
\sigma_{\textrm{dc}}=\frac{4e^{2}}{h}\frac{1}{16n_{i}}\left[1+\left(\frac{8v_{F}^{2}\hbar^{2}}{V_{0}t_{\perp}}\right)^{2}\right]k_{F}^{2}.\label{eq:SigmaDC_delta_BLG}\end{equation}
 Although for the bilayer both the FBA and the exact calculation results
in the conductivity being proportional to $k_{F}^{2}$, we should
note that in the former case the conductivity is proportional to the
strength of the potential, and therefore the FBA cannot be trusted
in the regime of strong potentials, and the agreement on the $k_{F}$-dependence
of the two approaches is fortuitous. (We remark that the limitations
of the FBA for a description of electronic transport are not exclusive
to short-range scatterers and can also be found in Coulomb scatterers.\cite{coulombnovikov})

The results of the present and previous sections confirm the intuitive
idea that delta potentials and hardwall (hard-disk) boundary conditions
originate the same dependence of $\sigma_{\textrm{dc}}$ on the Fermi
momentum. Remarkably, letting $V_{0}\rightarrow\infty$ in Eqs.~(\ref{eq:sigmaDC_delta_SLG})
and (\ref{eq:SigmaDC_delta_BLG}), give precisely Eqs.~(\ref{eq_sigma_graphene})
and (\ref{eq_sigma_BL}), respectively, and hence the two models are
equivalent with regard to strong short-range potentials.

\subsection{Quantum corrections near the neutrality point\label{sub:KPM}}

The Boltzmann approach beyond the FBA provides a good description
of the effect of strong short-range scatterers on the transport properties
of graphene at finite carrier densities (and for not too large concentrations
of resonant impurities).\cite{Klos2010} However, near the neutrality
point quantum interference effects become important and a fully quantum
calculation is needed to assess dc-transport. (For recent reviews
on the importance of quantum effects in the transport properties of
graphene see Refs.~\onlinecite{PeresRMP2010},~\onlinecite{EduardoReview}.)
In what follows, we present large-lattice, tight-binding numerical
calculations in the low-density regime and finite (high) impurity
concentration limit $n_{\textrm{ad}}\sim1\%$, where quantum corrections
due to multi-scattering events cannot be ignored.

\begin{figure}
\begin{centering}
\includegraphics[width=8cm]{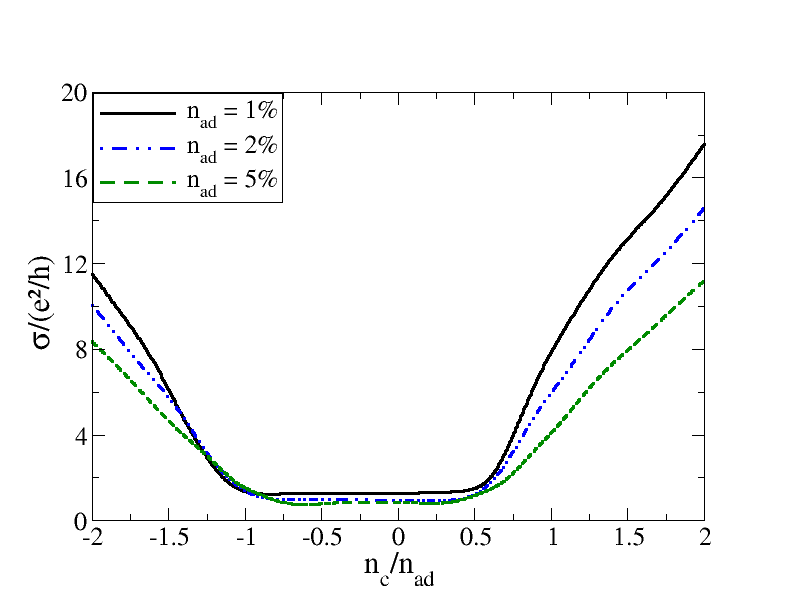} 
\par\end{centering}

\caption{\label{fig:KPM_COND_SLG}(Color online) Conductivity as function of
the normalized carrier density $n_{\textrm{c}}/n_{\textrm{ad}}$ for
a monolayer honeycomb lattice with $N=1000\times1000$ for different
concentrations of adsorbed atoms (periodic boundary conditions and
ten realizations of disorder were taken). The tight-binding parameters
read $V_{\textrm{ad}}=2t$ and $\epsilon_{\textrm{ad}}=-0.0625t$
. }

\end{figure}

\emph{Monolayer graphene}---We start by extending the monolayer tight-binding
Hamiltonian {[}Eqs.~(\ref{eq_tbhamilt}) and (\ref{eq:rs_contribution_ham}){]}
to include a finite number $N_{\textrm{ad}}$ of adsorbed atoms of
the same species, binding to carbons placed at (random) positions
$\{\mathbf{s}_{i}\}\;(i=1,...,N_{\textrm{ad}})$, \begin{eqnarray}
\hat{H}_{\textrm{tb}} & = & -t\sum_{n,\bm{\delta}_{i}}\vert\bm{R}_{n},A\rangle\langle\bm{R}_{n}+\bm{\delta}_{i},B\vert+\textrm{H.c.}\nonumber \\
 &  & +\sum_{i=1}^{N_{\textrm{ad}}}\left[V_{\textrm{ad}}\vert\mathbf{s}_{i},\textrm{ad}\rangle\langle\mathbf{s}_{i},C_{i}\vert+\textrm{H.c.}\right.\nonumber \\
 &  & \left.+\epsilon_{\textrm{ad}}\vert\mathbf{s}_{i},\textrm{ad}\rangle\langle\mathbf{s}_{i},\textrm{ad}\vert\right]\:,\label{eq:tight-binding-Hamilt}\end{eqnarray}
 where $C_{i}=A$($B$) for adatoms binding to carbon atoms in the
$A$($B$) sublattice. The Kubo formula for the zero-temperature dc-conductivity
tensor reads\cite{Mahan} \begin{equation}
\sigma_{ab}(E)=\frac{2\pi\hbar e^{2}}{A}\textrm{Tr}\left[\hat{v}_{a}\,\delta(E-\hat{H}_{\textrm{tb}})\,\hat{v}_{b}\,\delta(E-\hat{H}_{\textrm{tb}})\right],\label{eq:Kubo_formula}\end{equation}
 where $\hat{v}_{a(b)}$ is the $a(b)$th component of the velocity
operator (defined through the Heisenberg equation of motion for the
position coordinate) and $A$ stands for area of graphene.

We evaluate the longitudinal component of the conductivity $\sigma_{xx}$
employing a KPM: details of the calculation are given elsewhere.\cite{KPM_to_be_published}
The KPM amounts to approximate functions defined in bounded intervals
by a truncated sum over polynomials with optimized weights.\cite{Weisse-Review}
To illustrate the change in the transport properties near the neutrality
point, we simulate mesoscopic-size square sheets of graphene with
$N=10^{6}$ carbon sites. An adequate polynomial expansion of Eq.~(\ref{eq:Kubo_formula})
allows us to perform the simulations with modest computational resources.

We found that the expansion of Eq.~(\ref{eq:Kubo_formula}) in Chebyshev
polynomials of the first kind converges for concentrations of resonant
impurities, $n_{\text{ad}}=N_{\textrm{ad}}/N$, above a critical value
$n_{\textrm{ad}}^{*}$ of about $1\%$ (for $N=10^{6}$). We interpret
this result as an indication that for $n_{\text{ad}}<n_{\text{ad}}^{*}$,
electronic carriers are in the ballistic regime. (Recall that only
in diffusive or localized regimes can a thermodynamic conductivity
be defined.) The values $n_{\textrm{ad}}\ge n_{\textrm{ad}}^{*}$
correspond to concentrations of short-range scatterers several orders
of magnitude larger than what is found in typical laboratory environments
(about $10^{-3}$\%; see previous sections and Ref. \onlinecite{Dpeak})
but can, in principle, be reached via hydrogenation of graphene on
SiO$_{2}$.\cite{Katoch2010} The critical value $n_{\text{ad}}^{*}$
likely indicates the onset of diffusive behavior, $l\leq L$, where
$l$ is the mean free path and $L$ denotes the lattice linear size.
Thus, in principle be lowered by increasing $L$.

Figure~\ref{fig:KPM_COND_SLG} shows results for conductivity as
function of the carrier density; the latter was obtained by integration
of the density of states $\rho(E)$ (shown in Fig.~\ref{fig:KPM_DOS}),
according to \begin{equation}
n_{\textrm{c}}(E_{F})=\left(g_{s}/D\right)\int_{0}^{E_{F}}\rho(E)dE,\label{eq:carrier_density}\end{equation}
 where $D=N+N_{\textrm{ad}}$ is the total dimension of the problem.
The most peculiar feature in Fig.~\ref{fig:KPM_COND_SLG} is the
plateau of finite conductivity, due to the formation of a low-energy
impurity band (Fig.~\ref{fig:KPM_DOS}, top), a particular case of
disorder-enhanced conductivity.\cite{Titovsingle,Caio,Ozone}

The dc conductivity at the neutrality point differs significantly
from calculations based on Boltzmann kinetic theory. (1) The conductivity
saturates at a low carrier density to a finite value $\sigma_{\textrm{min}}>0$
around $e^{2}/h$ (the precise value depends on $n_{\textrm{ad}}$
and sample size), in accordance with theoretical predictions.\cite{Mirlin2010}
The width of the saturation is roughly proportional to the density
of adatoms in the probed range of impurity concentration $n_{\textrm{ad}}\le5$\%
(a similar behavior was first reported using a self-consistent approximation
to the Green's function of the electrons in the presence of a strong
disordered potential\cite{nmrPRB06} and recently reported in Ref.~\onlinecite{wehlingII}).
(2) The conductivity (for a fixed carrier density or energy) is not
proportional to $1/n_{\textrm{ad}}$. {[}In fact, a careful inspection
of the KPM conductivity data discloses that the latter observation
extends to higher carrier densities: (resonant) adsorbate-limited
transport in small samples of graphene displays a rich behavior until
full diffusive transport is reached.{]} Both fact 1 and fact 2 clearly
indicate that we are operating outside the applicability of the Boltzmann
approach.

\begin{figure}
\begin{centering}
\includegraphics[width=8cm]{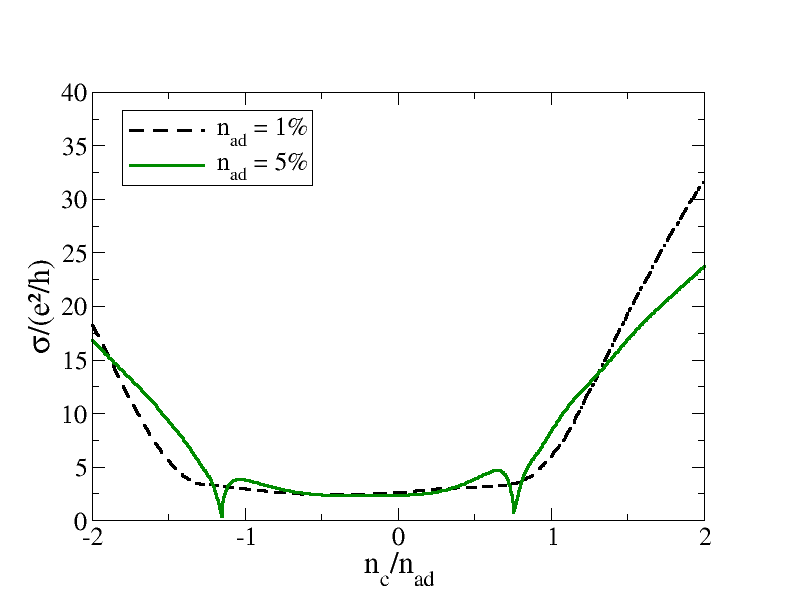} 
\par\end{centering}

\begin{centering}
\includegraphics[width=8cm]{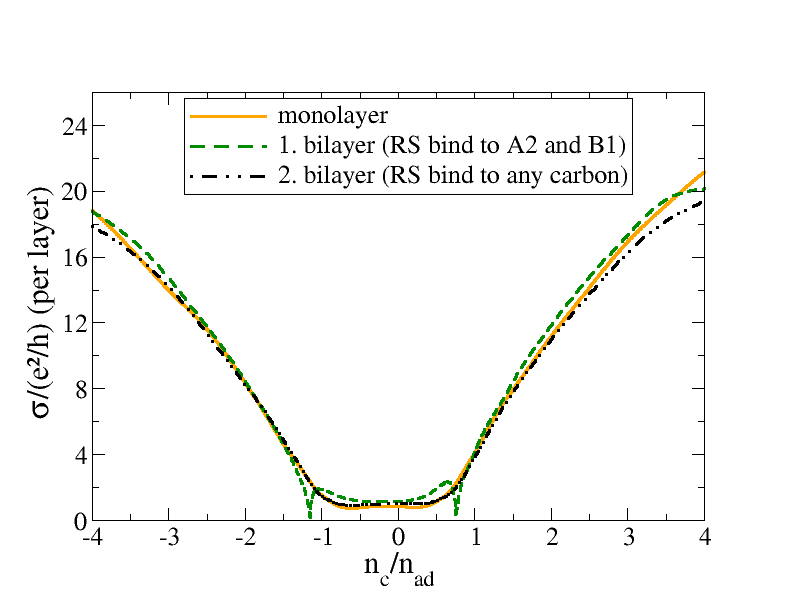} 
\par\end{centering}

\caption{\label{fig:KPM_COND_BLG}(Color online) Top: Conductivity as a function
of the normalized carrier density $n_{\textrm{c}}/n_{\textrm{ad}}$
for a bilayer honeycomb lattice with $N=2\times1000\times1000$ for
different concentrations of adsorbed atoms (periodic boundary conditions
and 10 realizations of disorder were taken). The tight-binding parameters
read $V_{\textrm{ad}}=2t$, $\epsilon_{\textrm{ad}}=-0.0625t$, and
$t_{\perp}=0.2t$. Bottom: Comparison of the conductivity (per layer)
of monolayer and bilayer graphene (with $n_{\textrm{ad}}=5\%$ in
both cases). Two bilayer curves are shown corresponding to different
arrangements of resonant scatterers (RS) as discussed in Sec.~(\ref{sec_resonant}):
(1) adsorbates binding only to carbons $A_{2}$ and $B_{1}$, and
(2) adsorbates forming bonds with carbons in any sublattice. The former
situation leads to a supression of the plateau near the edges.}

\end{figure}

Our results, in general, agree well with those reported in Ref.~\onlinecite{wehlingII}
for larger lattices (where $N$ of the order of $10^{8}$ was used).
Notwithstanding, we point out some differences concerning the plateau
of conductivity minimum: we observe neither peaks within the conductivity
plateau (including for $n_{\textrm{ad}}$ = 5\%) nor a plateau's width
of $2\times n_{\textrm{ad}}$, as claimed in that work. This could
be due to the different methods and system sizes used (although in
simulations with a larger lattice, we found no evidence of both effects).

A comment about intervalley scattering in our simulations is in order;
Anderson localization induced by intervalley scattering will become
experimentally relevant and prevent conductivity saturation only for
either very strong disorder (i.e., high defect densities) or exceedingly
large samples at very low temperatures. In contrast, our results,
and those in Ref. \onlinecite{wehlingII}, show no evidence for
localization even for relatively high amounts of resonant disorder.
This suggests that the localization length due to resonant scatterers
is far larger than that obtained for an on-site Anderson model, hence
allowing for conductivity-induced disorder, $\sigma_{0}>0$, in typical-size
graphene samples.

\emph{Bilayer graphene} -- The tight-binding Hamiltonian for bilayer
graphene with resonant impurities reads\begin{eqnarray}
\hat{H}_{\textrm{tb}}^{\textrm{(BLG)}} & = & \hat{H}_{\textrm{tb}}^{\textrm{(L=1,2)}}+t_{\perp}\sum_{n,\bm{\delta}_{i}}\left(\vert\bm{R}_{n},A_{1}\rangle\langle\bm{R}_{n},B_{2}\vert+\textrm{H.c.}\right)\nonumber \\
 &  & +\sum_{\underset{(L=1,2)}{i=1}}^{N_{\textrm{ad}}}\left[V_{\textrm{ad}}\vert\mathbf{s}_{i}^{L},\textrm{ad}\rangle\langle\mathbf{s}_{i}^{L},C_{L}\vert+\textrm{H.c.}\right.\nonumber \\
 &  & \left.+\epsilon_{\textrm{ad}}\vert\mathbf{s}_{i}^{L},\textrm{ad}\rangle\langle\mathbf{s}_{i}^{L},\textrm{ad}\vert\right]\:,\label{eq:tight-binding-Hamilt-BLG}\end{eqnarray}
 where $\hat{H}_{\textrm{tb}}^{\textrm{(L=1,2)}}$ is the Hamiltonian
of two uncoupled layers ($L=1,2$) {[}see Eq.~(\ref{eq:tight-binding-Hamilt}){]},
the term with $t_{\perp}$ describes electronic interlayer hopping,
and the third term accounts for adsorbates binding to carbons in random
positions $\{\mathbf{s}_{i}^{L}\}$ in both layers. We choose $C_{1}(C_{2})=A_{2}(B_{1})$
to guarantee that adsorbates bind only to carbons with coordination
number $z=3$. {[}The transport properties when adatoms bind to carbons
in both sublattices are similar to those of monolayer graphene; see
Sec.~\ref{sec_resonant} and Fig.~\ref{fig:KPM_COND_BLG} (bottom).{]}

The conductivity of bilayer graphene follows from evaluating the Kubo
formula {[}Eq.~(\ref{eq:Kubo_formula}){]} with $\hat{H}_{\textrm{tb}}\rightarrow\hat{H}_{\textrm{tb}}^{\textrm{(BLG)}}$.
The KPM results (summarized in Fig.~\ref{fig:KPM_COND_BLG}) resemble
those obtained previously for monolayer graphene (Fig.~\ref{fig:KPM_COND_SLG}),
but with important differences. (1) The formation of the impurity
band leads to a conductivity minimum about twice the value found for
monolayer graphene {[}$\sigma_{\textrm{min}}\approx e^{2}/h$ (per
layer){]}. {[}This fact has been predicted before by coherent potential
approximation calculations of disorder in multilayer graphene.\cite{nilsson1,nilsson2}.
See Eqs.~(11) and (53) in Refs.~\onlinecite{nilsson1},~and \onlinecite{nilsson2},
respectively.{]} (2) For a high impurity concentration, $n_{\textrm{ad}}=5\%$,
the conducitvity is strongly suppressed before actually forming the
plateau; this curious effect is rooted in the opening of a gap in
bilayer graphene spectrum, due to the adsorbed species, uncoupling
the midgap region from higher energy states (see Fig.~\ref{fig:KPM_DOS},
bottom, and Fig.~\ref{fig:KPM_DOS_b}). In this case, we can then
speak of a {}``conduction gap.''

The bottom panel in Fig.~\ref{fig:KPM_COND_BLG} compares the conductivity
of monolayer and bilayer graphene for $n_{\textrm{ad}}=5\%$: away
from the plateau, as carriers have energies similar to or higher than
the interlayer coupling, $t_{\perp}$, we expect these systems to
have comparable conductivities (per graphene layer). Our results indeed
confirm the latter point, although we found that for a very high carrier
density, $|n_{\textrm{c}}|\gtrsim20\%$, the conductivity of both
systems cannot be compared reliably within our KPM approach: increasing
the carrier density up to such values originates carrier energies
close to the Von Hove singularities, and strong (spurious) numerical
oscillations in the KPM expansion cannot be avoided. In addition,
these oscillations behave differently in both systems (in particular,
because bilayer graphene has four such singularities), making any
comparison difficult. This is the reason why we have presented the
conductivity for low carrier densities, which also coincides with
the most relevant experimental regime.

We finish this section by noting that vacancy-induced disorder leads
to effects similar to those reported here, a fact satisfactorily explained
by the model of strong short-range scatterers presented in Sec.~\ref{sec_resonant}.
For vacancies, though, the strong conductivity electron-hole asymmetry
(caused by the offset resonant peaks) will not be present.

\section{Scattering in a biased bilayer graphene\label{sec_scattBBL}}

\begin{figure}[ht]
 \centering{} \includegraphics[width=7.5cm]{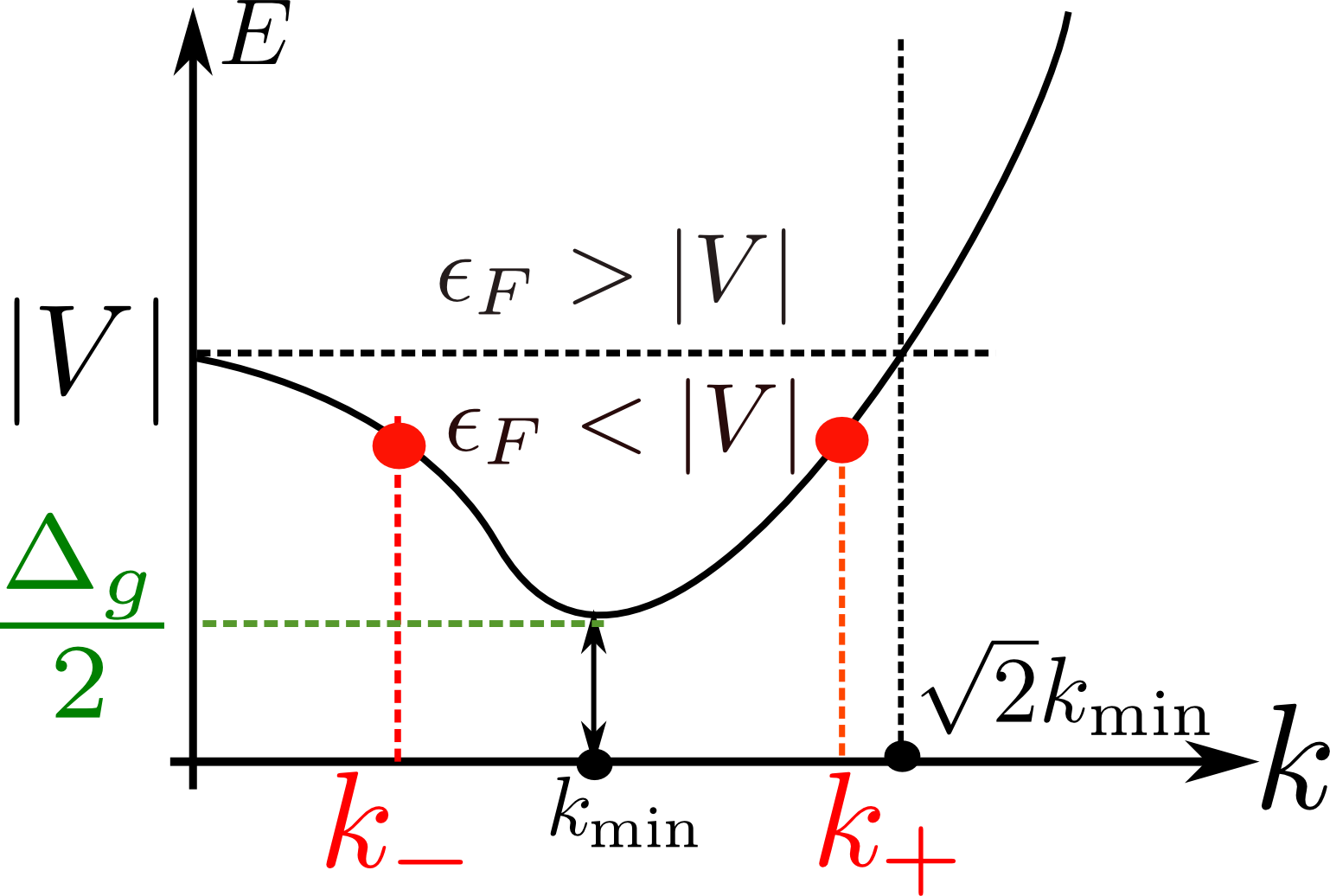} \caption{\label{fig_BBL_spectrum}(Color online) Energy spectrum of a biased
graphene bilayer. Several quantities defined in the text are depicted,
and $E_{F}$ stands for the Fermi energy. Information on the two regimes
$E_{F}\gtrless|V|$ is included. Full circles represent degenerate
states with energy $E=E(k_{+})=E(k_{-})$, a fact that will have to
be taken into account when establishing a scattering theory.}

\end{figure}

When $V\ne0$, electrons in a graphene bilayer are described by Eq.~(\ref{eq_HBLVneZ}).
In this case, the energy spectrum develops a Mexican hat form, as
represented in Fig.~\ref{fig_BBL_spectrum}, and the spectrum opens-up
a gap. When the energy of the electrons is lower than $|V|$, the
Fermi surface becomes a ring around the Dirac point, with an inner,
$k_{-}$, and an outer, $k_{+}$, Fermi radius in momentum space.\cite{GuineaRing,StauberRing}

Therefore, for $E<|V|$, we have two degenerate states with different
momentum values. As we show below, the description of scattering in
these two regimes, $E\gtrless|V|$, is necessarily different.

The regular eigenstates of Hamiltonian in Eq.~(\ref{eq_HBLVneZ})
in polar coordinates are given by \begin{equation}
\Psi_{m}(r,\theta)=\frac{1}{\sqrt{A}}\left[\begin{array}{c}
a_{k}J_{m}(kr)\\
\mp b_{k}J_{m+2}(kr)e^{2i\theta}\end{array}\right]e^{im\theta}\,,\label{eq_Eigenstates_BBL}\end{equation}
 to which corresponds the eigenvalues \begin{equation}
E(k)=\pm\sqrt{V^{2}(1-\epsilon_{k}/t_{\perp})^{2}+\epsilon_{k}^{2}}\,,\label{eq_spectrum_BBL}\end{equation}
 where $\epsilon_{k}=v_{F}^{2}\hbar^{2}k^{2}/t_{\perp}$ is the energy
of electrons in bilayer graphene for $V=0$, and the coefficients
$a_{k}$ and $b_{k}$ read \begin{eqnarray}
a_{k} & = & \sqrt{\frac{1}{2}}[1+V(1-\epsilon_{k}/t_{\perp})/E]^{1/2}\,,\\
b_{k} & = & \sqrt{\frac{1}{2}}[1-V(1-\epsilon_{k}/t_{\perp})/E]^{1/2}\,.\end{eqnarray}
 Additionally, the relation $a_{k}^{2}b_{k}^{2}=\epsilon_{k}^{2}/(4E^{2})$
holds. %

The density probability flux $J_{\ell}$ is given by Eq.~(\ref{eq_JellBL}),
plus an additional term $J_{\ell}^{V}$, reading \begin{equation}
J_{\ell}^{V}=2V\frac{v_{F}^{2}\hbar}{t_{\perp}^{2}}\textrm{Im}\Psi^{\dagger}\hat{J}_{\ell}^{V}\Psi\,,\label{eq_JellBL_V_correction}\end{equation}
 where the operator $\hat{J}_{\ell}^{V}$ is given by \begin{equation}
\hat{J}_{\ell}^{V}=\left[\begin{array}{cc}
-\partial_{\ell} & 0\\
0\  & \partial_{\ell}\end{array}\right]\,.\end{equation}

Throughout, we consider that electronic carriers have positive energy
$E>0$ (the other case follows immediately). Let us establish here
some useful relations for later use. The energy gap $\Delta_{g}$
is determined by \begin{equation}
\Delta_{g}=2E(k_{\textrm{min}})=2|V|t_{\perp}\left(V^{2}+t_{\perp}^{2}\right)^{-1/2}\,,\end{equation}
 where $k_{\textrm{min}}$ is defined in Eq.~(\ref{eq_kminus_kplus}).
Given a state with energy $E$, the two momentum values are obtained
from the inversion of the energy spectrum, Eq.~(\ref{eq_spectrum_BBL}),
and are given by the positive roots of the equation \begin{equation}
\frac{\epsilon_{k}}{t_{\perp}}=\frac{\Delta_{g}^{2}}{4t_{\perp}^{2}}\left[1\pm f(E)\right]\,,\label{eq_momentum_roots}\end{equation}
 with $f(E)=\sqrt{1-(1+t_{\perp}^{2}/V^{2})(1-E^{2}/V^{2})}$. From
Eq.~(\ref{eq_momentum_roots}) we see that for $E<|V|$ the two roots
are real, corresponding to two propagating states, whereas for $E>|V|$,
only one root is real, corresponding to a single propagating state;
this is consistent with the dispersion depicted in Fig.~\ref{fig_BBL_spectrum}.
In the latter regime, the imaginary root is essential to fulfill the
scattering boundary conditions, as in the case discussed in Sec.~\ref{subsecBL}.
For energy $E=|V|$, we are at the boundary between the two regimes
introduced above: $E\gtrless|V|$. In this case, the scattering descriptions
below and above $E=|V|$ must provide the same answer. For $E=|V|$
we have $k_{-}=0$ and $k_{+}=\Delta_{g}/(\sqrt{2}v_{F}\hbar)=\sqrt{2}k_{\textrm{min}}$;
for $E<|V|$ we have a simple relation between $k_{-}$ and $k_{+}$,
reading \begin{equation}
k_{-}=\sqrt{2k_{\textrm{min}}^{2}-k_{+}^{2}}\hspace{0.5cm}{\rm \textrm{and}}\hspace{0.5cm}k_{\textrm{min}}=\frac{\Delta_{g}}{2v_{F}\hbar}\,.\label{eq_kminus_kplus}\end{equation}
 The radial velocity of the electrons at $k_{-}$ and $k_{+}$ is
given by\begin{equation}
v_{r}(k_{\pm})=\frac{2v_{F}^{2}\hbar}{t_{\perp}}\frac{V^{2}f(E)}{t_{\perp}E}\left(\pm k_{\pm}\right).\label{eq:velocity}\end{equation}
 Clearly, the state with momentum $k_{-}$ has a negative velocity;
the scattering formalism has to take this aspect into account.

Because the regimes $E>|V|$ and $E<|V|$ are distinct, in the sense
that the latter case contains two degenerate propagating states, we
develop the scattering theory separately for both cases.

\subsection{The $E>|V|$ regime \label{sec_scattBBL_E_bigger_V}}

For $E>|V|$, the two momenta are $k_{+}=k$ and $k_{-}=i\sqrt{k_{+}^{2}-2k_{\textrm{min}}^{2}}=i\kappa$.
The latter value originates an evanescent wave at the boundary of
the potential. As in the case in Sec. \ref{subsecBL}, it is simple
to show that a wave function of the form

\begin{equation}
\Psi(\mathbf{r})\simeq\frac{1}{\sqrt{A}}\left(\begin{array}{c}
a_{k_{x}}\\
b_{k_{x}}\end{array}\right)e^{ik_{+}x}+\frac{1}{\sqrt{A}}\left(\begin{array}{lc}
a_{k}\\
b_{k}e^{2i\theta}\end{array}\right)f(\theta)\frac{e^{ik_{+}r}}{\sqrt{r}}\,,\label{eq_scattBBL_E_bigger_V}\end{equation}
 represents an incoming plane wave of momentum $\mathbf{k}_{i}=(k_{+},0)\equiv(k,0)$
and a scattered cylindrical wave of momentum $\mathbf{k}_{f}=k_{+}(\cos\theta,\sin\theta)$.
Note that relative to the case of the unbiased bilayer case, Eq.~(\ref{eq_s-wave_limit_E_greater_V})
differs in the presence of the amplitudes $a_{k}$ and $b_{k}$. The
scattered radial flux has the usual form $J_{r}=v_{r}(k)\vert f(\theta)\vert^{2}/r$,
from which the differential cross section follows as $\sigma(\theta)=\vert f(\theta)\vert^{2}$.
As in Sec.~\ref{sec_boltzmann_GBL}, we seek a wave function in the
form of a superposition of Bessel functions of different kinds, which
in the present case can be written as \begin{eqnarray}
\Psi_{m}(r,\theta) & = & A_{1}^{m}\left[\begin{array}{rc}
a_{k}J_{m}(kr)\\
-b_{k}e^{2i\theta}J_{m+2}(kr)\end{array}\right]\nonumber \\
 & + & A_{2}^{m}\left[\begin{array}{rc}
a_{k}Y_{m}(kr)\\
-b_{k}e^{2i\theta}Y_{m+2}(kr)\end{array}\right]\nonumber \\
 & + & A_{3}^{m}\left[\begin{array}{rc}
a_{\kappa}K_{m}(\kappa r)\\
-b_{i\kappa}e^{2i\theta}K_{m+2}(\kappa r)\end{array}\right]\,.\label{eq_waveBBL_Well}\end{eqnarray}
 The ratio $A_{m}^{2}/A_{1}^{m}$ reads \begin{equation}
\frac{A_{2}^{m}}{A_{1}^{m}}=\frac{a_{k}b_{i\kappa}J_{m}(kR)K_{m+2}(\kappa R)-b_{k}a_{i\kappa}J_{m+2}(kR)K_{m}(\kappa R)}{b_{k}a_{i\kappa}K_{m}(\kappa R)Y_{m+2}(kR)-a_{k}b_{i\kappa}K_{m+2}(\kappa R)Y_{m}(kR)}\,.\label{eq_ratioA1A2BBL_energy_large_V}\end{equation}
 Combining Eqs.~(\ref{eq_tangdel0A1A2BL}) and (\ref{eq_ratioA1A2BBL_energy_large_V}),
the equation for the phase shift $\delta_{m}$ follows at once. Indeed,
the expression for the dc conductivity of electrons with Fermi momentum
$k_{+}$ is similar to Eq.~(\ref{eq_sigmaDC}), reading\begin{equation}
\sigma_{\textrm{dc}}=\frac{4e^{2}}{h}\frac{k_{+}^{2}}{4n_{i}\Lambda(k_{+})}.\,\label{eq:sigmaDC_E.gt.V}\end{equation}
 In the regime $k_{+}\gg\sqrt{2}k_{\textrm{min}}$, we have $\kappa\approx k_{+}=k$,
$a_{k}\approx a_{i\kappa}$, and $b_{k}\approx b_{i\kappa}$, and
therefore the phase shifts given by Eq.~(\ref{eq_ratioA1A2BL}) and
(\ref{eq_ratioA1A2BBL_energy_large_V}) are essentially identical;
that is, we have \begin{equation}
\delta_{0}\rightarrow\frac{\pi}{2}\,,\hspace{0.3cm}(k_{+}\gg\sqrt{2}k_{\textrm{min}})\,.\label{eq_Phase_Shift_BBL_high_density}\end{equation}
 As a consequence of Eq.~(\ref{eq_Phase_Shift_BBL_high_density}),
the conductivity is essentially linear in $V_{g}$ at a high electronic
density.

When the gate voltage is reduced, bringing the Fermi energy close
to $V$, we have $\kappa\rightarrow0$, but $k_{+}\gtrsim\sqrt{2}k_{\textrm{min}}$
is finite. In this case, we have \begin{equation}
\frac{A_{2}^{m}}{A_{1}^{m}}\rightarrow-\frac{J_{m+2}(kR)}{Y_{m+2}(kR)}\,,\label{eq_Phase_Shift_BBL_low_density}\end{equation}
 and considering that $kR\lesssim1$, the $s$ wave phase shift tends
to \begin{equation}
\delta_{0}\rightarrow-\frac{\pi}{8}\left(k_{\textrm{min}}R\right)^{4}\hspace{0.3cm}\mbox{for}\ k_{+}\rightarrow\sqrt{2}k_{\textrm{min}}\,.\label{eq_s-wave_limit_E_greater_V}\end{equation}
 The bias potential acts differently on electron and hole carriers
{[}see Eq.~(\ref{eq_HBLVneZ}){]}, with the effect that the symmetry
relation between phase shifts changes to $\delta_{m}(E,V)=\delta_{-m-2}(-E,V)$.
Also, the phase shifts for negative energy carriers (holes) must equal
the phase shifts for positive energy carriers (electrons) if the sign
of $V$ is reversed.

\begin{figure}[ht]
 \centering{} \includegraphics[width=8cm]{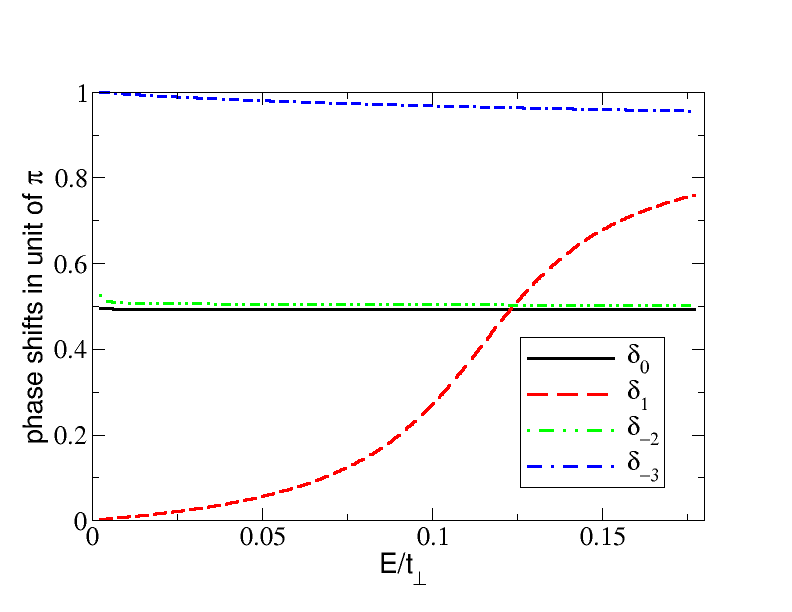} \caption{\label{fig_phase_shift_bias} (Color online) Dependence of nonzero
phase shifts $\delta_{m}$ on $E$, for the biased bilayer graphene
with $R=a_{0}$ for low electrostatic potential $V=4\times10^{-3}t_{\perp}$.
The energy range here excludes the interval $[4,5[$ ($10^{-3}t_{\perp}$)
for which the energy begins to fall within the {}``Mexican hat''
(Fig.~\ref{fig_BBL_spectrum}). In the vicinity of $E=|V|$, we have
$k_{+}\rightarrow\sqrt{2}k_{\textrm{min}}$ and the $s$ wave phase
shift $\delta_{0}$ for electrons (or $\delta_{-2}$ for holes) drops
quickly to the value indicated in Eq.~(\ref{eq_s-wave_limit_E_greater_V}).}

\end{figure}

The dependence of the phase shifts on the gates voltage (that is,
on both $k$ and $V$) is now more involved. Figure~\ref{fig_phase_shift_bias}
shows the non-zero phase shifts for electrons for the particular case
of weak interlayer potential $V$. Similarly to the unbiased bilayer
($V=0$) there are four (non-zero) phase shifts, however, as stressed
above, the presence of the interlayer potential lifts the degeneracy
observed in Fig.~\ref{fig_phase_shift_01}; in particular, for $|V|>0$
the phase shifts with $m=-1$ and $m=-3$ differ very much (except
for energies very close to $V$). On the contrary, the phase shifts
$\delta_{0}$ and $\delta_{-2}$ just differ significantly close to
the vicinity of $E=V$, where the systems approaches the {}``Mexican
hat.''

\subsection{The $E<|V|$ regime \label{sec_scattBBL_E_smaller_V}}

As discussed at the beginning of Sec.~\ref{sec_scattBBL}, in the
case $E<|V|$ there are two degenerate propagating states, characterized
by $k_{-}$ and $k_{+}$. Thus, the matrix element of the potential
between these two states is finite, and an incoming particle with
a well-defined momentum ($k_{-}$ or $k_{+}$) will be scattered in
a superposition of both momenta. This fact requires the modification
of the scattering formalism introduced above.

In what follows, we develop the scattering formalism assuming that
the incoming electron has momentum $k_{+}$; the case where the incoming
electron has momentum $k_{-}$ follows immediately, and only the final
results are given.

We start by assuming that the total wave function in the presence
of the potential, at large distances from it, has the asymptotic form
\begin{eqnarray}
\Psi(\mathbf{r}) & \simeq & \frac{1}{\sqrt{A}}\left[\begin{array}{c}
a_{k_{x}}\\
b_{k_{x}}\end{array}\right]e^{ik_{+}x}+\frac{1}{\sqrt{A}}\left[\begin{array}{lc}
a_{k_{+}}\\
b_{k_{+}}e^{2i\theta}\end{array}\right]f_{++}(\theta)\frac{e^{ik_{+}r}}{\sqrt{r}}\nonumber \\
 & + & \frac{1}{\sqrt{A}}\left[\begin{array}{lc}
a_{k_{-}}\\
b_{k_{-}}e^{2i\theta}\end{array}\right]f_{+-}(\theta)\frac{e^{-ik_{-}r}}{\sqrt{r}}\,,\label{eq_scattBBL_E_smaller_V}\end{eqnarray}
 where $f_{++}(\theta)$ represents the scattering amplitude considering
that the outgoing electron has the same momentum, $k_{+}$, as the
incoming one, and $f_{+-}(\theta)$ represents the scattering amplitude
considering that the outgoing electron changed its momentum to $k_{-}$.
Let us stress again that $E(k_{-})=E(k_{+})$. Since the velocity
of the state with momentum $k_{-}$ is negative, the sign of the argument
in the exponential of associated cylindrical wave function has to
be negative, since these states represent particles propagating backward
in time (a positive sign gives a radial incoming flux). The fluxes
associated with the first, second, and third terms on the right-hand
side of Eq.~(\ref{eq_scattBBL_E_smaller_V}) read \begin{equation}
J_{x}^{+}=v_{x}(k_{+})\,,\,,\label{eq:flux_Jx}\end{equation}
 \begin{equation}
J_{r}^{+}=v_{r}(k_{+})\vert f_{++}(\theta)\vert^{2}r^{-1}\,,\label{eq:flux_Jr_plus}\end{equation}
 and \begin{equation}
J_{r}^{-}=-v_{r}(k_{-})\vert f_{+-}(\theta)\vert^{2}r^{-1}\,,\label{eq:flux_Jr_minus}\end{equation}
 respectively, from which follows the existence of two scattering
cross sections, defined as \begin{equation}
\sigma_{++}(\theta)=\vert f_{++}(\theta)\vert^{2}\hspace{0.2cm}\textrm{and}\hspace{0.2cm}\sigma_{+-}(\theta)=-\frac{v_{r}(k_{-})}{v_{r}(k_{+})}\vert f_{+-}(\theta)\vert^{2}\,.\label{eq:diff_cross_sec1}\end{equation}
 Both cross sections must enter in the relaxation time needed to compute
the dc conductivity.

We now assume that a partial wave in the angular momentum basis of
the total wave function has, at large distances from the potential,
the form

\begin{eqnarray}
\Psi_{m}(r,\theta) & \simeq & \left[\begin{array}{lc}
a_{k_{+}}\\
b_{k_{+}}e^{2i\theta}\end{array}\right]\frac{e^{-i(k_{+}r-\lambda_{m}-m\theta)}}{\sqrt{2\pi Ak_{+}r}}\nonumber \\
 & + & \eta_{m,++}e^{2i\delta_{m,++}}\left[\begin{array}{lc}
a_{k_{+}}\\
b_{k_{+}}e^{2i\theta}\end{array}\right]\frac{e^{i(k_{+}r-\lambda_{m}+m\theta)}}{\sqrt{2\pi Ak_{+}r}}\nonumber \\
 & + & \eta_{m,+-}\left[\begin{array}{lc}
a_{k_{-}}\\
b_{k_{-}}e^{2i\theta}\end{array}\right]\frac{e^{-i(k_{-}r-\lambda_{m}-m\theta)}}{\sqrt{2\pi Ak_{-}r}}\,,\label{eq_scatt_angular_E_smaller_V}\end{eqnarray}
 where $\delta_{m,++}$ is the phase shift of the partial wave $m$,
$0<\eta_{m,++}<1$ is a real number accounting for the transfer of
probability flux to the outgoing momentum channel $k_{-}$, and $0<\vert\eta_{m,+-}\vert^{2}<1$.
Conservation of the radial flux for each partial wave $m$ imposes
\begin{equation}
\eta_{m,++}^{2}+\vert\eta_{m,+-}\vert^{2}=1\,.\label{eq_flux_conserv}\end{equation}
 Summing over $m$, according to Eq.~(\ref{eq_psi_dirac_angular}),
we obtain $\Psi(\mathbf{r})$ in the form given by Eq.~(\ref{eq_scattBBL_E_smaller_V}),
with the scattering amplitudes defined as \begin{eqnarray}
f_{++} & = & \frac{1}{\sqrt{2\pi ik_{+}}}\sum_{m}\left(\eta_{m,++}e^{2i\delta_{m,++}}-1\right)e^{im\theta}\,,\label{eq:f++}\\
f_{+-} & = & \frac{1}{\sqrt{2\pi ik_{-}}}\sum_{m}\eta_{m,+-}e^{im\theta}\,.\label{eq:f+-}\end{eqnarray}

As in Sec.~\ref{subsecBL}, we write the exact partial wave of the
full scattering problem, for $r>R$, as \begin{eqnarray}
\Psi_{m}(r,\theta) & = & A_{1}^{m}\left[\begin{array}{lc}
a_{k_{+}}H_{m}^{(2)}(k_{+}r)\\
-b_{k_{+}}H_{m+2}^{(2)}(k_{+}r)e^{2i\theta}\end{array}\right]\nonumber \\
 & + & A_{2}^{m}\left[\begin{array}{lc}
a_{k_{+}}H_{m}^{(1)}(k_{+}r)\\
-b_{k_{+}}H_{m+2}^{(1)}(k_{+}r)e^{2i\theta}\end{array}\right]\nonumber \\
 & + & A_{3}^{m}\left[\begin{array}{lc}
a_{k_{-}}H_{m}^{(2)}(k_{-}r)\\
-b_{k_{-}}H_{m+2}^{(2)}(k_{-}r)e^{2i\theta}\end{array}\right]\,.\label{eq_exact_scatt_angular_E_smaller_V}\end{eqnarray}
 Expanding Eq.~(\ref{eq_exact_scatt_angular_E_smaller_V}) for large
$r$ and comparing it with Eq.~(\ref{eq_scatt_angular_E_smaller_V}),
we see that \begin{equation}
\frac{A_{2}^{m}}{A_{1}^{m}}=\eta_{m,++}e^{2i\delta_{m,++}}\,,\hspace{0.2cm}\frac{A_{3}^{m}}{A_{1}^{m}}=\eta_{m,+-}\,.\end{equation}
Calculation of the differential cross section requires the determination
of $\eta_{m,++}$, $\eta_{m,+-}$, and $\delta_{m,++}$. In the limit
$V_{0}\rightarrow\infty$, the boundary condition is $\Psi_{m}(r=R)=0$,
leading to

\begin{widetext} \begin{eqnarray}
\frac{A_{2}^{m}}{A_{1}^{m}} & = & \eta_{m,++}e^{2i\delta_{m,++}}=\frac{a_{k_{+}}b_{k_{-}}H_{m}^{(2)}(k_{+}R)H_{m+2}^{(2)}(k_{-}R)-b_{k_{+}}a_{k_{-}}H_{m+2}^{(2)}(k_{+}R)H_{m}^{(2)}(k_{-}R)}{b_{k_{+}}a_{k_{-}}H_{m}^{(2)}(k_{-}R)H_{m+2}^{(1)}(k_{+}R)-a_{k_{+}}b_{k_{-}}H_{m+2}^{(2)}(k_{-}R)H_{m}^{(1)}(k_{+}R)}\,,\label{eq_etapp}\\
\frac{A_{3}^{m}}{A_{1}^{m}} & = & \eta_{m,+-}=-a_{k_{+}}b_{k_{+}}\frac{H_{m+2}^{(1)}(k_{+}R)H_{m}^{(2)}(k_{+}R)-H_{m}^{(1)}(k_{+}R)H_{m+2}^{(2)}(k_{+}R)}{b_{k_{+}}a_{k_{-}}H_{m}^{(2)}(k_{-}R)H_{m+2}^{(1)}(k_{+}R)-a_{k_{+}}b_{k_{-}}H_{m+2}^{(2)}(k_{-}R)H_{m}^{(1)}(k_{+}R)}\,.\label{eq_ratioA1A2BBL_energy_smaller_V}\end{eqnarray}
 \end{widetext}

Although not immediately obvious, the parameters $\eta_{m,++}$ and
$\eta_{m,+-}$, as given by Eqs.~(\ref{eq_etapp}) and (\ref{eq_ratioA1A2BBL_energy_smaller_V}),
obey the flux conservation relation in Eq.~(\ref{eq_flux_conserv}).
When the Fermi energy, $E_{F}$, approaches the energy $E=|V|$ from
below, we have $k_{-}\rightarrow0$. In this limit, we find \begin{eqnarray}
\eta_{0,++}e^{2i\delta_{0,++}} & \rightarrow & -\frac{H_{2}^{(2)}(k_{+}R)}{H_{2}^{(1)}(k_{+}R)}\,,\label{eq_limit_below}\\
\eta_{m,+-} & \rightarrow & 0\,,\end{eqnarray}
 as it should. Since $k_{+}R\lesssim1$, it follows from Eq.~(\ref{eq_limit_below})
that \begin{equation}
\delta_{0,++}\rightarrow-\frac{\pi}{32}\left(k_{\textrm{+}}R\right)^{4}\,,\end{equation}
 which, for $\delta_{0,++}$, gives the same result found in Eq.~(\ref{eq_s-wave_limit_E_greater_V}).

The above results hold for an incoming electron with momentum $k_{+}$;
when the electron has momentum $k_{-}$ we have the cross sections:
\begin{equation}
\sigma_{--}(\theta)=\vert f_{--}(\theta)\vert^{2}\hspace{0.2cm}\textrm{and}\hspace{0.2cm}\sigma_{-+}(\theta)=-\frac{v_{r}(k_{+})}{v_{r}(k_{-})}\vert f_{-+}(\theta)\vert^{2}\,,\label{eq:diff_cross_sec2}\end{equation}
 whose amplitudes are given by the right-hand side of Eqs.~(\ref{eq:f++})
and (\ref{eq:f+-}), respectively, upon interchanging $k_{+}$ with
$k_{-}$.

\subsection{dc conductivity of a biased bilayer graphene \label{sec_DC_BBL}}

As discussed in Sec.~\ref{sec_boltzmann_GBL}, calculation of the
dc conductivity requires the computation of the exact phase shifts.
We start by studying the behavior of the $s$ wave phase shift as
a function of the Fermi momentum for a biased graphene bilayer.

\begin{figure}[ht]
 \centering{} \includegraphics[width=8cm]{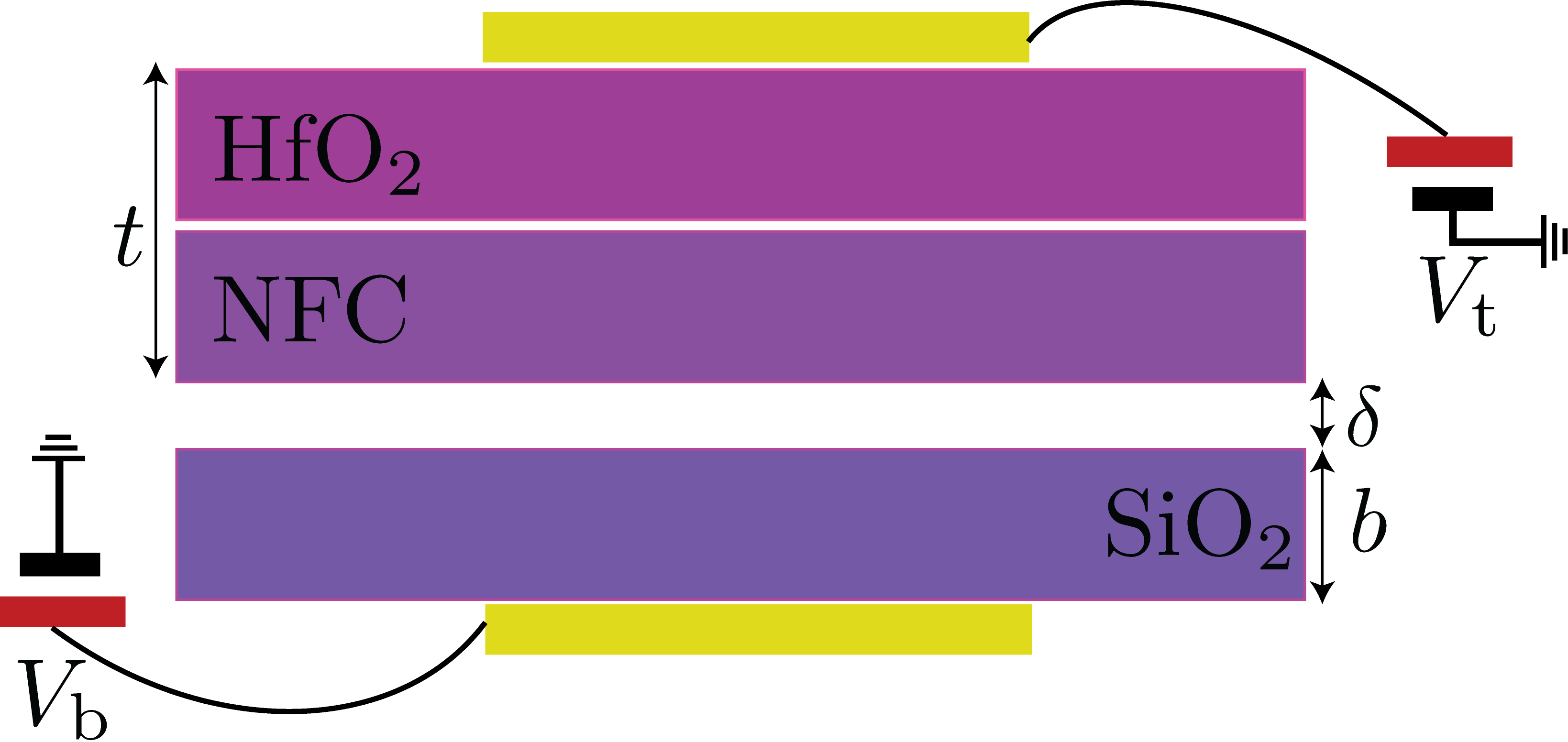} \caption{\label{fig_BBL_gates}(Color online) Capacitor geometry for dual-gate
transistor. \cite{avourisonoff} The figure is self-explanatory. Values
of the several quantities are: $\delta=3.4$ \AA{}, $b=300$ nm, and
$t=20$ nm. $V_{\textrm{t}}$ and $V_{\textrm{b}}$ stand for the
top and bottom gate potentials, respectively.}

\end{figure}

In the biased bilayer, the ability of independently tuning the electronic
density and the value of the gap $\Delta_{g}$ requires the use of
two gates, a bottom and a top gates, as shown in Fig.~\ref{fig_BBL_gates}.
The electric field in the top-gate dielectric is ($e>0$) \begin{equation}
E_{\textrm{t}}=\frac{en_{\textrm{t}}}{\epsilon_{\textrm{t}}\epsilon_{0}}\,,\label{eq_Etop}\end{equation}
 and that in the bottom-gate dielectric is \begin{equation}
E_{\textrm{b}}=\frac{en_{\textrm{b}}}{\epsilon_{\mathbf{\textrm{b}}}\epsilon_{0}}\,,\label{eq_Ebottom}\end{equation}
where $n_{\textrm{t}}$ and $n_{\textrm{b}}$ are the electronic density
in the top and bottom gate, respectively, and $\epsilon_{\textrm{t}}$
and $\epsilon_{\mathbf{\textrm{b}}}$ are the relative permittivity
of the top- and bottom-gate dielectric, respectively. Charge neutrality
requires that the total amount of charge accumulated in the bilayer
is $-en=-e(n_{\textrm{t}}+n_{\textrm{b}})$. The electrostatic potential
difference between the top gate and the bilayer is $V_{\textrm{t}}=tE_{\textrm{t}}$,
whereas between the bottom gate and the bilayer it is $V_{\textrm{b}}=bE_{\textrm{b}}$.
It follows from Eqs.~(\ref{eq_Etop}) and (\ref{eq_Ebottom}) that
\begin{equation}
V_{\textrm{b}}=b\frac{en_{b}}{\epsilon_{\mathbf{\textrm{b}}}\epsilon_{0}}=\frac{ben}{\epsilon_{\mathbf{\textrm{b}}}\epsilon_{0}}-\frac{b\epsilon_{\textrm{t}}}{t\epsilon_{\mathbf{\textrm{b}}}}V_{\textrm{t}}\,.\label{eq_Vbottom}\end{equation}
 Inverting Eq.~(\ref{eq_Vbottom}), the total electronic density
in the bilayer is given by \begin{equation}
n=V_{\textrm{b}}\frac{\epsilon_{\textrm{b}}\epsilon_{0}}{be}+\frac{\epsilon_{0}\epsilon_{\textrm{t}}}{et}V_{\textrm{t}}\,.\label{eq:carrier_density_bias}\end{equation}
 When $n$ is positive, the bilayer is doped with electrons; when
$n$ is negative the system is doped with holes. Finally, the electrostatic
potential difference between the two graphene layers in the bilayer
is given by \begin{equation}
\Delta V=(E_{\textrm{b}}-E_{\textrm{t}})\delta=\frac{ne\delta}{\epsilon_{\mathbf{\textrm{b}}}\epsilon_{0}}-\left(\frac{\epsilon_{\textrm{t}}}{\epsilon_{\mathbf{\textrm{b}}}}+1\right)\frac{\delta}{t}V_{\textrm{t}}\,,\label{eq:electrostatic_potential_bias}\end{equation}
 where $\delta=3.4$ \AA{} is the interlayer distance (we are ignoring
screening effects,\cite{McCann2,castro,castro_2010} which are not
important for small $V_{{\rm t}}$). The variable $V$ introduced
in Eq.~(\ref{eq_HBLfull}) relates to $\Delta V$ as $2V=\Delta V$.
Taking typical values for dual-gate bilayer transistors, \cite{avourisonoff}
we have: $\epsilon_{\textrm{SiO}_{2}}=3.9$, $\epsilon_{\textrm{HfO}_{2}}=25$,
$\epsilon_{\textrm{NFC}}=2.4$, $b=300$ nm, and $t=20$ nm (both
dielectrics, HfO$_{2}$ and NFC, have about the same width). The relative
permittivity of $\epsilon_{t}$ is \begin{equation}
\epsilon_{\textrm{t}}=\frac{2\epsilon_{\textrm{HfO}_{2}}\epsilon_{\textrm{NFC}}}{\epsilon_{\textrm{HfO}_{2}}+\epsilon_{\textrm{NFC}}}\,.\end{equation}
 In working devices,\cite{avourisonoff} we have $|V_{\textrm{b}}|\lesssim70$
V and $|V_{\textrm{t}}|\lesssim4$ V.

The calculation of dc conductivity follows, as before, from Boltzmann's
transport theory. In the regime $E>|V|$, $\sigma_{\textrm{dc}}$
is still given by Eq.~(\ref{eq_sigmaDC}), but with the phase shifts
determined from Eq.~(\ref{eq_ratioA1A2BBL_energy_large_V}). When
$E<|V|$, there are two scattering channels and this implies that
the resulting formula for $\sigma_{\textrm{dc}}$ differs somewhat
from that given in Eq.~(\ref{eq_sigmaDC}), reading \begin{equation}
\sigma_{\textrm{dc}}=\frac{4e^{2}}{h}\frac{1}{2}\left[\frac{k_{+}}{n_{i}\sigma(k_{+})}+\frac{k_{-}}{n_{i}\sigma(k_{-})}\right]\,,\label{eq:sigma_dc_E.le.V}\end{equation}
 where $\sigma(k_{\pm})$ is defined as \begin{equation}
\sigma(k_{\pm})=\int_{0}^{2\pi}[\sigma_{\pm,+}(\theta)+\sigma_{\pm,-}(\theta)](1-\cos\theta)\,.\label{eq:transp_cross_sec}\end{equation}
 Inserting the expressions for the differential cross sections {[}Eqs.~(\ref{eq:diff_cross_sec1})
and (\ref{eq:diff_cross_sec2}){]} and performing the integral yields

\begin{widetext}\begin{align}
\sigma(k_{\pm})= & \frac{1}{k_{\pm}}\textrm{Re}\sum_{m}\left[\left|\eta_{m,\pm\pm}e^{2i\delta_{m,\pm\pm}}-1\right|^{2}-\left(\eta_{m,\pm\pm}e^{2i\delta_{m,\pm\pm}}-1\right)\left(\eta_{m+1,\pm\pm}e^{-2i\delta_{m+1,\pm\pm}}-1\right)\right.\nonumber \\
 & \left.-\frac{k_{\pm}}{k_{\mp}}\frac{v_{r}(k_{\mp})}{v_{r}(k_{\pm})}\left(|\eta_{m,\pm\mp}|^{2}-\eta_{m,\pm\mp}\eta_{m+1,\pm\mp}^{*}\right)\right]\,.\label{eq:transp_cross_sec-1}\end{align}
 \end{widetext}

The formulas for $\sigma_{--}$ and $\sigma_{-+}$ are identical and
thus are not presented. The dc conductivity follows from the determination
of the Fermi momentum, given the carrier density in the bilayer, which
in turn depends on both gates as given by Eq.~(\ref{eq:carrier_density_bias}).
The relation $k_{F}^{2}=\pi n$ (valid for various two-dimensional
systems) must be adapted to take into account the degeneracy of the
spectrum (Fig.~\ref{fig_BBL_spectrum}) and reads, \begin{equation}
k_{F}^{+}=\sqrt{\frac{\pi}{2}n+k_{\textrm{min}}^{2}}\,,\label{eq:k_fermi_E.le.V}\end{equation}
 and the other propagating state $k_{F}^{-}$ relates to $k_{F}^{+}$
according to Eq.~(\ref{eq_kminus_kplus}).

\begin{figure}
\begin{centering}
\includegraphics[width=8cm]{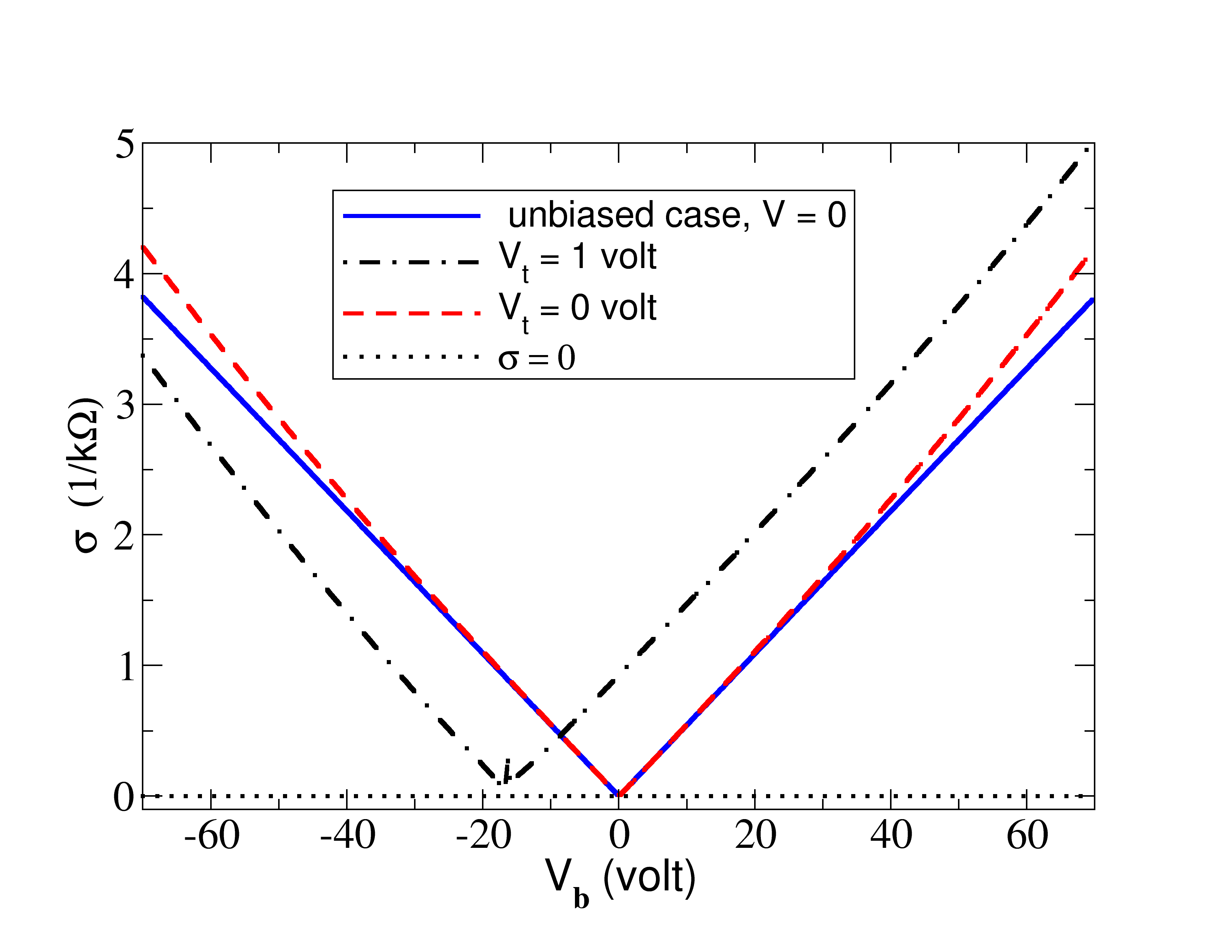} 
\par\end{centering}

\caption{\label{fig:BBL_cond}(Color online) Dependence of the biased bilayer
dc conductivity on the back-gate potential $V_{\textrm{b}}$ (values
of $V_{\textrm{t}}$ are indicated). The solid line shows the dc conductivity
for the unbiased case $V=0$, for comparison. }

\end{figure}

Figure~\ref{fig:BBL_cond} shows the dc conductivity as function
of the back gate for fixed values of $V_{\textrm{t}}$. As the back
gate $V_{\textrm{b}}$ is varied, the gap $\Delta_{g}$ and the Fermi
energy change; for a small window of width $\sim1$V around $V_{\textrm{b}}\simeq-17$
V the system moves into the regime $E<|V|$ and expression in Eq.~(\ref{eq:k_fermi_E.le.V})
must be used to determine the carriers energy. In this energy regime,
$k_{F}$ is bounded according to $\sqrt{2}k_{\textrm{min}}\ge k_{F}\ge k_{\textrm{min}}$
and hence the value $k_{F}=0$ is forbidden; as a consequence, and
at odds with the unbiased bilayer, the minimum conductivity is not
exactly zero, having a value of $\sigma_{\textrm{min}}\simeq3e^{2}/h$
for $V_{\textrm{t}}=1$ V.

\section{Conclusion}

\label{sec_conclusion}

In the early studies of transport in graphene, charged impurities
located in the substrate seemed to explain the measured conductivity.
Recent experiments suggest other possibility though.\cite{Dpeak,Katoch2010}
While there is a consensus that electron and hole puddles, induced
by charged impurities, dominate the landscape near the neutrality
point, away from this point, adsorbed hydrocarbons, at the surface
of graphene, may be the limiting factor in dc-transport.

In the present paper, we established an intuitive theoretical picture
of scattering due to resonant scattering originated by adatoms. Although
resonant scatterers have been studied before (first in Refs. \onlinecite{robinson,
stauberBZ} and, more recently, in Ref. \onlinecite{wehlingII}),
we have established the first coherent picture of resonant-scattering
limited dc transport valid for both monolayer and bilayer graphene.

Section \ref{sec_resonant} reviews the electronic structure of monolayer
graphene and presents, for the first time, the density of states of
bilayer graphene with resonant contaminants. Despite the distinct
electronic structure of pristine monolayer and bilayer graphene, this
section shows that resonant adatoms lead to the same effect in both
systems: the emergence of resonant peaks in the vicinity of the Dirac
point, a situation reminiscent of vacancy-induced disorder.\cite{vitorpaco,Castro2010}
Using a simple tight-binding toy model, resonant adatoms are seen
to be reliably mimicked by a particular class of short-range scatterers,
that is, those having an intrinsic energy much higher than typical
graphene energies. This fact motivates the subsequent study of dc
transport using strong short-range potentials in a continuum formulation
(Secs. \ref{sec_boltzmann_GSL} and \ref{sec_boltzmann_GBL}).

Section \ref{DCcond} shows that the typical dependence of conductivity
with the electronic density in the monolayer (sublinear dependence)
and bilayer (linear dependence) systems can be explained assuming
resonant scatterers alone. The comparison with experimental data bears
out the agreement with dc transport experiments performed in exfoliated
few-layer graphene films, hence providing further strength to the
resonant-scatterer hypothesis. To justify the robustness of a continuum-model
approach based on strong short-range scatterers as prototypes of real
resonant adsorbates, we have calculated the semiclassical conductivity
due to two types of strong local potentials (hard-disk and delta-potential),
finding perfect agreement between the two methods (partial wave analysis
and Lippmann-Schwinger equation, respectively) and tested the validity
of the long wavelength limit (on the basis of the continuum formulation)
against numerical lattice calculation using a $T$ matrix approach
(Sec. \ref{sub:tmatrix}).

Section \ref{sub:delta} demonstrates the incorrectness of the widely
used FBA within the semiclassical (Boltzmann) approach, in the context
of short-range disorder, and the need to compute the electronic scattering
amplitudes as accurately as possible, hence, clarifying an issue overlooked
in the graphene literature. Section \ref{sub:KPM} presents the Kubo
dc conductivity evaluated numerically with a KPM; from this calcution,
the breakdown of the semiclassical picture close to neutrality, in
the regime of a high concentraton of impurities, is clearly observed.
Here, the case of bilayer graphene is addressed for the first time,
with the results showing that a {}``conduction gap'' takes place
for selective adsorbate bonding, due to a strong supression of the
conducitivy in the surroundings of the resonant impurity band.

Finally, due to its importance for technological applications, scattering
in the bilayer graphene with a gap in the sectrum is studied in Sec.
\ref{sec_scattBBL}, by extending the well-established partial wave
method (Sec. \ref{partialwave}) to describe scattering in the biased
bilayer graphene. Such a scattering theory has never been developed
before (to the best of our knowledge) and can be easily adapted to
tackle other physical scenarios requiring the need for computing scattering
amplitudes when the energy dispersion relation is degenerate.

We are confident that our results help to elucidate the electronic
transport properties of this remarkable two-dimensional material. 

Note added: After submission of this work for publication, we become
aware of a paper\cite{Yuan2} which also discusses the effect of resonant
scatterers on the dc conductivity of single-layer and bilayer graphene,
with results that are consistent with ours.

\section*{Acknowledgements}

A.F. acknowledges FCT Grant No. SFRH/BPD/65600/2009. E.R.M. acknowledges
partial financial support by NSF Grant No. DMR 1006230. A.H.C.N. acknowledges
Grant No. DE-FG02-08ER46512. Discussions with A. K. Geim are acknowledged.

\section{Appendix}

In this appendix, starting from the low-energy continuum theory, we
derive the nonperturbative semiclassical dc conductivity of monolayer
and bilayer graphene with short-range scatterers. This calculation
requires the solution of the two-dimensional scattering problem, where
a massless fermion with incident momentum $\mathbf{p=\hbar k}$ is
brought to interaction with an impurity. We model the potential of
the impurity by a delta function, $V_{d}=V_{0}\delta(\mathbf{r})$.
Following standard methods, the formal solution of $(H_{0}+V_{d}-E)\Psi_{\mathbf{k}}=0$
can be written as,\begin{equation}
\Psi_{\mathbf{k}}=\phi_{\mathbf{k}}+\hat{G}_{0}\hat{V}_{d}\Psi_{\mathbf{k}}\,,\label{eq:A9}\end{equation}
where $H_{0}$ is the low-energy Hamiltonian of graphene; $\phi_{\mathbf{k}}$
is the solution of the free problem $(H_{0}-E)\phi_{\mathbf{k}}=0$
and describes the state of the incident particles. Here, $H_{0}$
refers to the Hamiltonian obtained from expansion of the graphene
dispersion around the $\mathbf{K}$ point (the calculation involving
the remaining valley is equivalent). The resolvent is given by $\hat{G}_{0}=1/(E+i0^{+}-H_{0})$,
and the energy includes a small positive imaginary part $i0^{+}$.
The spinor $\phi_{\mathbf{k}}(\mathbf{r})\equiv\langle\mathbf{r}|\phi_{\mathbf{k}}\rangle$
has the form \cite{rmp}, \begin{equation}
\phi_{\mathbf{k}}(\mathbf{r})=u_{\mathbf{k}}^{(\lambda)}e^{i\mathbf{k}\cdot\mathbf{r}}\,,\label{eq:A9b}\end{equation}
 with,\begin{equation}
u_{\mathbf{k}}^{(\lambda)}=\frac{1}{\sqrt{2A}}\left(\begin{array}{c}
1\\
se^{i\lambda\theta_{\mathbf{k}}}\end{array}\right)\,.\label{eq:A9b2}\end{equation}
The Berry phase is $\varphi_{B}\equiv\pi\lambda$ and equals $\pi$
for monolayer graphene, whereas for bilayer graphene its value is
$2\pi$ {[}compare with Eq.~(\ref{eq_spinora}){]}. The second component
of $u_{\mathbf{k}}$ includes the sign $s=\pm$ of the electronic
carrier charge and $\theta_{\mathbf{k}}\equiv\arctan(k_{y}/k_{x})$.
Switching Eq.~(\ref{eq:A9}) to the position representation, we obtain
the Lippmann-Schwinger equation,\begin{equation}
\Psi_{\mathbf{k}}(\mathbf{r})=\phi_{\mathbf{k}}(\mathbf{r})+\int d^{2}\mathbf{r}^{\prime}G_{0}(\mathbf{r}-\mathbf{r}^{\prime})V(\mathbf{r}^{\prime})\Psi_{\mathbf{k}}(\mathbf{r^{\prime}})\,.\label{eq:A10}\end{equation}
 In the latter equation, $G_{0}(\mathbf{r}-\mathbf{r}^{\prime})=\langle\mathbf{r}|\left(E+i0^{+}-H_{0}\right)^{-1}|\mathbf{r}^{\prime}\rangle$
is the Green function of the problem. Monolayer graphene has $H_{0}=\hbar v_{F}\boldsymbol{\sigma}\cdot\hat{\mathbf{p}}$
and the Fourier transform of the Green function obeys, \begin{equation}
\left(E+i0^{+}-\boldsymbol{\sigma}\cdot\mathbf{p}\right)G_{0}(\mathbf{p})=1\,,\label{eq:A11}\end{equation}
 where $G_{0}(\mathbf{p})=\int d\mathbf{r}\exp\left(-i\mathbf{p}\cdot\mathbf{r}\right)G_{0}(\mathbf{r})$
(notice that to simplify notation, we have set $\hbar=1$ and $v_{F}=1$).
Inverting the $2\times2$ matrix on the left-hand side of Eq.~(\ref{eq:A11}),
we arrive at\begin{eqnarray}
G_{0}(\mathbf{p}) & = & g_{1}(\mathbf{p})\left(E+\boldsymbol{\sigma}\cdot\mathbf{p}\right)\,,\label{eq:A12}\\
g_{1}(\mathbf{p}) & = & 1/\left[E^{2}-p^{2}+i0^{+}\right].\end{eqnarray}

The calculations for $E>0$ and $E<0$ are similar and to be specific
we focus on the former situation. Indeed, the inclusion of a small
imaginary part from positive values $i0^{+}$ amounts to consider
outgoing waves (see below). We write $E=k$ and evaluate the Green
function in the real space representation,\begin{align}
G_{0}(\mathbf{r}-\mathbf{r}^{\prime}) & =\frac{1}{4\pi^{2}}\left(E-i\boldsymbol{\sigma}\cdot\mathbf{\nabla}\right)\int d^{2}\mathbf{p}e^{i\mathbf{p}\cdot(\mathbf{r}-\mathbf{r}^{\prime})}g_{1}(\mathbf{p})\label{eq:A13a}\\
 & =-\frac{i}{4}\left(k-i\boldsymbol{\sigma}\cdot\mathbf{\nabla}\right)H_{0}^{(1)}\left(k|\mathbf{r}-\mathbf{r}^{\prime}|\right)\,,\label{eq:A13b}\end{align}
 where $H_{n}^{(1)}\left(k|\mathbf{r}-\mathbf{r}^{\prime}|\right)$
is the Hankel function of the first kind of order $n$, whose asymptotic
form is that of outgoing cylindrical waves {[}see Eq.~(\ref{eq:A17a}){]}.
The Hankel function obeys $\partial_{x}H_{0}^{(1)}(x)+H_{1}^{(1)}(x)=0$;
hence, \begin{equation}
\boldsymbol{\sigma}\cdot\mathbf{\nabla}H_{0}^{(1)}(k|\mathbf{r}-\mathbf{r}^{\prime}|)=-kH_{1}^{(1)}(k|\mathbf{r}-\mathbf{r}^{\prime}|)\sigma_{\theta}\,,\label{eq:A14}\end{equation}
 where we have introduced the matrix, \begin{equation}
\sigma_{\theta}\equiv\left(\begin{array}{cc}
0 & e^{-i\theta}\\
e^{i\theta} & 0\end{array}\right)\,,\label{eq:A15}\end{equation}
 and the angle $\theta\equiv\theta(\mathbf{r},\mathbf{r}^{\prime})$
is defined by $\left(\mathbf{r}-\mathbf{r}^{\prime}\right)/|\mathbf{r}-\mathbf{r}^{\prime}|=\left(\cos\theta,\sin\theta\right)^{T}$.
Combining Eqs.~(\ref{eq:A14}) and (\ref{eq:A13b}), we have, at
once,\begin{equation}
G_{0}(\mathbf{r}-\mathbf{r}^{\prime})=-\frac{ik}{4}\left[H_{0}^{(1)}(k|\mathbf{r}-\mathbf{r}^{\prime}|)+i\sigma_{\theta}H_{1}^{(1)}(k|\mathbf{r}-\mathbf{r}^{\prime}|)\right]\,.\label{eq:A15b}\end{equation}

The derivation of the Green function of bilayer graphene follows identical
steps. We write the free Hamiltonian as $H_{0}=-(v_{F}^{2}\hbar^{2}/t_{\perp})\boldsymbol{\sigma}\cdot\mathbf{D}$,
with $\mathbf{D}=(\partial_{x}^{2}-\partial_{y}^{2},\,2\partial_{x}\partial_{y})^{T}$.
As before, we set $\hbar$ and $v_{F}$ temporarily equal to the unit;
the Fourier transform of bilayer Green function reads \begin{equation}
G_{0}(\mathbf{p})=g_{2}(\mathbf{p})\left[E+\gamma\mathbf{\boldsymbol{\sigma}}\cdot\tilde{\mathbf{D}}(\mathbf{p})\right]\,,\label{eq:Propagator}\end{equation}
 where $\gamma\equiv1/t_{\perp}$, $E=\gamma k^{2}$, $\tilde{\mathbf{D}}(\mathbf{p})=(p_{x}^{2}-p_{y}^{2},\,2p_{x}p_{y})^{T}$,
and \begin{equation}
g_{2}(\mathbf{p})=\frac{1}{2E}\left[\frac{1}{E-\gamma p^{2}+i0^{+}}+\frac{1}{E+\gamma p^{2}+i0^{+}}\right]\,.\label{eq:Scalar_propagator}\end{equation}
 Since $H_{0}$ is quadratic in momentum operators, $g_{2}$ resembles
a non relativistic propagator. Again, we focus on the case of electrons
($\gamma>0$),

\begin{equation}
G_{0}(\mathbf{r}-\mathbf{r}^{\prime})=\frac{\gamma}{4\pi^{2}}\left(k^{2}-\boldsymbol{\sigma}\cdot\mathbf{D}\right)\int d^{2}\mathbf{p}e^{i\mathbf{p}\cdot(\mathbf{r}-\mathbf{r}^{\prime})}g_{2}(\mathbf{p})\,.\label{eq:aux3}\end{equation}
 The contribution to the integrand of Eq.~(\ref{eq:aux3}) with poles
in the real axis can be simplified using, \begin{equation}
\frac{1}{k^{2}-p^{2}+i0^{+}}=\frac{i\pi}{2k}\left[\delta(p+k)+\delta(p-k)\right]+\textrm{P.V.}\frac{1}{k^{2}-p^{2}}\,.\label{eq:PV_relation}\end{equation}
 Performing the integral in Eq.~(\ref{eq:aux3}) yields,\begin{eqnarray}
G_{0}(\mathbf{r}-\mathbf{r}^{\prime}) & = & \frac{1}{8\gamma k^{2}}\left(k^{2}-\sigma\cdot\mathbf{D}\right)\left[-iH_{0}^{(1)}\left(k|\mathbf{r}-\mathbf{r}^{\prime}|\right)\right..\nonumber \\
 &  & \left.+\frac{2}{\pi}K_{0}\left(k|\mathbf{r}-\mathbf{r}^{\prime}|\right)\right]\,.\label{eq:Propagator_Exact}\end{eqnarray}
 The first term in brackets describes scattered waves in two dimensions,
whereas the modified Bessel function $K_{0}$ describes evanescent
waves (recall that $k\rightarrow ik$ is a solution of $H_{0}$ with
the same energy). For short-range potentials the main contribution
to the scattering amplitude comes from evaluating Eq.~(\ref{eq:A10})
within the region where $|\mathbf{r}-\mathbf{r}^{\prime}|\gg1$ and
hence $K_{0}$ will not contribute (see later).

In what follows, we compute the nonperturbative scattering amplitude
for monolayer and bilayer graphene, which will be needed for the calculation
of the dc conductivity in these systems.

\subsection*{A1. Nonperturbative amplitude for monolayer graphene }

Inserting the expression of the potential $V_{d}=V_{0}\delta(\mathbf{r})$
in the Lippmann-Schwinger equation {[}Eq.~(\ref{eq:A10}){]} and
performing the spatial integration results in,\begin{equation}
\Psi_{\mathbf{k}}(\mathbf{r})=\phi_{\mathbf{k}}(\mathbf{r})+V_{0}G_{0}(\mathbf{r})\Psi_{\mathbf{k}}(0)\,,\label{eq:-3}\end{equation}
which is ill defined because putting $\mathbf{r}=0$ yields a divergence,
namely, $\Psi_{\mathbf{k}}(0)=\phi_{\mathbf{k}}(0)+(\infty)$. This
stems from the singularity of $G_{0}(\mathbf{r})$ {[}Eq.~(\ref{eq:A15b}){]}
at the origin $\mathbf{r}=0$, a common situation in field theories.
The only way of curing this divergence is by means of renormalization.\cite{Weinberg}
Let us explicitly describe this procedure. The explicit expression
for $G_{0}(0)$ {[}see Eq.~(\ref{eq:A13a}){]} reads\begin{equation}
G_{0}(0)=\int\frac{d^{2}\mathbf{p}}{\left(2\pi\right)^{2}}\, g_{1}(|\mathbf{p}|)\left[E+\boldsymbol{\sigma}\cdot\mathbf{p}\right]\,,\label{eq:-4}\end{equation}
 Evaluating $G_{0}(\mathbf{r})$ at the origin and setting $E=k$
yields

\begin{equation}
G_{0}(0)\sim\int dp\frac{p}{k^{2}-p^{2}+i0^{+}}\,,\label{eq:A4-35-1}\end{equation}
 which is logarithmic divergent. To obtain a physical meaningful result,
a momentum cutoff, $p_{\textrm{max}}$, in the upper limit of the
integral must be considered. (Such procedure is justified because
graphene, being a solid, has an intrinsic energy cutoff of the order
of the bandwidth.) We thus have\begin{equation}
G_{0}(0)=\frac{k}{2\pi}\int_{0}^{p_{\textrm{max}}}dp\frac{p}{k^{2}-p^{2}+i0^{+}}\,.\label{eq:A4-35-2}\end{equation}
 This integral yields\begin{equation}
G_{0}(0)\cong\frac{k}{2\pi}\ln\left(kR\right)\,,\label{eq:A4-35-3}\end{equation}
 where we have assumed $k\ll p_{\textrm{max}}$ and $R\equiv1/p_{\textrm{max}}$
is a length scale of the order of $a_{0}$. Setting $\mathbf{r}=0$
in Eq.~(\ref{eq:-3}), using the latter result and solving for $\Psi_{\mathbf{k}}(0)$
gives \begin{eqnarray}
\Psi_{\mathbf{k}}(0) & = & \left[1-\frac{V_{0}}{2\pi}k\ln\left(kR\right)\right]^{-1}\phi_{\mathbf{k}}(0)\nonumber \\
 & = & \left[1-\frac{V_{0}}{2\pi}k\ln\left(kR\right)\right]^{-1}u_{\mathbf{k}}^{(1)}\,.\label{eq:A4-38}\end{eqnarray}
 To identify the scattered amplitude, we need the asymptotic form
of the Lippmann-Schwinger equation {[}Eq.~(\ref{eq:A10}){]}. For
short-range potentials the main contribution in Eq.~(\ref{eq:A10})
comes from the region where $|\mathbf{r}-\mathbf{r}^{\prime}|\gg1$.
Inserting the exact form of the propagator in space representation
{[}Eq.~(\ref{eq:A15b}){]} and using \begin{align}
H_{0}^{(1)}(k|\mathbf{r}-\mathbf{r}^{\prime}|) & \rightarrow\sqrt{\frac{2}{i\pi k|\mathbf{\mathbf{r}-\mathbf{r}^{\prime}}|}}e^{ik|\mathbf{r}-\mathbf{r}^{\prime}|}\,,\label{eq:A17a}\\
H_{1}^{(1)}(k|\mathbf{r}-\mathbf{r}^{\prime}|) & \rightarrow-i\sqrt{\frac{2}{i\pi k|\mathbf{\mathbf{r}-\mathbf{r}^{\prime}}|}}e^{ik|\mathbf{r}-\mathbf{r}^{\prime}|}\,,\label{eq:A17b}\end{align}
 leads to\begin{align}
\Psi_{\mathbf{k}}(\mathbf{r}) & =\phi_{\mathbf{k}}(\mathbf{r})\nonumber \\
 & -\sqrt{\frac{ik}{8\pi r}}e^{ikr}\int d^{2}\mathbf{r}^{\prime}e^{-i\mathbf{k}_{\textrm{out}}\cdot\mathbf{r}^{\prime}}\left(1+\sigma_{\theta}\right)V_{d}(\mathbf{r}^{\prime})\Psi_{\mathbf{k}}(\mathbf{r^{\prime}})\,,\label{eq:A18}\end{align}
 where we have approximated $|\mathbf{r}-\mathbf{r}^{\prime}|\simeq\mathbf{r}-\mathbf{r}\cdot\mathbf{r}^{\prime}/r$
and identified the wave vector at the point of observation, $\mathbf{k}_{\textrm{out}}\equiv k\mathbf{r}/r$.
The exact form of the spinor at the origin {[}Eq.~(\ref{eq:A4-38}){]}
allows us to find the explicit expression of $\Psi_{\mathbf{k}}(\mathbf{r})$;
letting $\tilde{\sigma}_{\theta}\equiv\sigma_{\theta}(\mathbf{r}^{\prime}=0)$,\begin{align}
\Psi_{\mathbf{k}}(\mathbf{r}) & =\phi_{\mathbf{k}}(\mathbf{r})\nonumber \\
 & -\frac{V_{0}}{1-\frac{V_{0}}{2\pi}k\ln\left(kR\right)}\sqrt{\frac{ik}{8\pi r}}e^{ikr}\left(1+\tilde{\sigma}_{\theta}\right)u_{\mathbf{k}}^{(1)}\,.\label{eq:A19}\end{align}

The action of $\left(1+\tilde{\sigma}_{\theta}\right)$ on the spinor
$u_{\mathbf{k}}^{(1)}$ yields the Berry phase term for scattering
in graphene; without loss of generality, we take the incident momentum
along the $x$ axis, $\mathbf{k}=(k,0)$, and thus \begin{eqnarray}
\left(1+\tilde{\sigma}_{\theta}\right)u_{\mathbf{k}}^{(1)} & = & \Xi_{B}(\theta)\frac{1}{\sqrt{2}A}\left(\begin{array}{c}
1\\
e^{i\theta}\end{array}\right)\nonumber \\
 & \equiv & \Xi_{B}(\theta)u_{\mathbf{k}_{\textrm{out}}}^{(1)}\,,\label{eq:A20}\end{eqnarray}
 where \begin{equation}
\Xi_{B}(\theta)=\left(1+e^{-i\theta}\right)\,,\label{eq:berry}\end{equation}
 and the scattering angle reads $\theta=\angle\left(\mathbf{k},\mathbf{k}_{\textrm{out}}\right)$
{[}recall Eq.~(\ref{eq:A15}) and comments therein{]}.

The wave function of the scattered particles is then\begin{equation}
\Psi_{\mathbf{k}}(\mathbf{r})=\phi_{\mathbf{k}}(\mathbf{r})+f(\theta)\frac{e^{ikr}}{\sqrt{r}}u_{\mathbf{k}_{\textrm{out}}}^{(1)}\,,\label{eq:A21}\end{equation}
 with the scattering amplitude reading,\begin{equation}
f(\theta)=-\frac{1}{\hbar v_{F}}\sqrt{\frac{ik}{8\pi}}\frac{V_{0}}{1-\frac{V_{0}}{2\pi\hbar v_{F}}k\ln\left(kR\right)}\Xi_{B}(\theta)\,,\label{eq:A22}\end{equation}
 where we have restored all the constants. (Note that here $V_{0}$
has units of {[}energy{]}$\times${[}length{]}$^{2}$; the relation
between $V_{0}$ and the effective impurity potential $V_{\textrm{eff}}$
in a lattice theory can be shown to be $V_{0}\sim A_{c}V_{\textrm{eff}}$,
where $A_{c}$ is the area of graphene's unit cell.) This result is
to be compared with the result from the FBA, which amounts to approximate
$\Psi_{\mathbf{k}}(\mathbf{r^{\prime}})$ by the unperturbed wave
function $\phi_{\mathbf{k}}(\mathbf{r}^{\prime})$ in Eq.~(\ref{eq:A18}):
\begin{equation}
f_{\textrm{Born}}(\theta)=-\frac{1}{\hbar v_{F}}\sqrt{\frac{ik}{8\pi}}V_{0}\Xi_{B}(\theta)\,.\label{eq:A23}\end{equation}
 The latter is only accurate in the limit of a very small $V_{0}$,
which is of limited interest. The nonperturbative result discloses
a singular momentum, $k_{\textrm{sing}}$,\begin{equation}
k_{\textrm{sing}}\ln\left(k_{\textrm{sing}}R\right)=\frac{2\pi\hbar v_{F}}{V_{0}}\,,\label{eq:A24}\end{equation}
 which corresponds to a bound state of our problem. More importantly,
the nonperturbative amplitude for $V_{0}\rightarrow\infty$ {[}recall
that resonant scatterers in graphene give origin to strong short-range
potentials, see Sec.~(\ref{sec_resonant}){]} reads\begin{equation}
f_{V\rightarrow\infty}(\theta)=\sqrt{\frac{i\pi}{2}}\frac{\Xi_{B}(\theta)}{\sqrt{k}\ln\left(kR\right)}\,,\label{eq:A25}\end{equation}
 which is the main result of the present section.

\subsection*{A2. Nonperturbative amplitude for bilayer graphene}

Calculation of the scattering amplitude for the bilayer graphene follows
as in Sec.~A1, albeit with the important difference that the Green
function does not diverge at the origin and thus no renormalization
procedure is needed this time. This explains why no regularization
length appears in the final result for the conductivity of bilayer
graphene. We now outline the derivation of this result.

The explicit expression for $G_{0}(0)$ {[}see Eq.~(\ref{eq:aux3}){]}
reads\begin{equation}
G_{0}(0)=\int\frac{d^{2}\mathbf{p}}{\left(2\pi\right)^{2}}\, g_{2}(\mathbf{p})\left[E+\gamma\mathbf{\boldsymbol{\sigma}}\cdot\tilde{\mathbf{D}}(\mathbf{p})\right]\,,\label{eq:BLG-1}\end{equation}
 which, setting $E=\gamma k^{2}$, can be simplified to \begin{equation}
G_{0}(0)=\gamma k^{2}\int\frac{d^{2}\mathbf{p}}{\left(2\pi\right)^{2}}\,\frac{1}{\left(\gamma k^{2}+i0^{+}\right)^{2}-\gamma^{2}p^{4}}\,.\label{eq:BLG-2}\end{equation}
 The above integral can be solved straightforwardly by contour integration;
the result is\begin{equation}
G_{0}(0)=-\frac{i}{8\gamma}\,.\label{eq:BLG-3}\end{equation}
 The amplitude of the wave function at the origin {[}Eq.~(\ref{eq:-3}){]}
therefore reads\begin{equation}
\Psi_{\mathbf{k}}(0)=\left[1+i\frac{V_{0}}{8\gamma}\right]^{-1}\phi_{\mathbf{k}}(0)\,.\label{eq:BLG-4}\end{equation}
To identify the scattered amplitude, we have to repeat the derivation
of the asymptotic form of the Lippmann-Schwinger equation {[}see Eqs.~(\ref{eq:A17a})--(\ref{eq:BLG-6}){]}.
The asymptotic form of the propagator can be calculated from Eq.~(\ref{eq:Propagator_Exact}),
\begin{equation}
G_{0}(\mathbf{r}-\mathbf{r}^{\prime})\rightarrow-\frac{1}{4\gamma}\sqrt{\frac{i}{2k\pi|\mathbf{r}-\mathbf{r}^{\prime}|}}\left(\begin{array}{cc}
1 & e^{-2i\theta(\mathbf{r},\mathbf{r}^{\prime})}\\
e^{2i\theta(\mathbf{r},\mathbf{r}^{\prime})} & 1\end{array}\right)e^{ik|\mathbf{r}-\mathbf{r}^{\prime}|}\,,\label{eq:BLG-5}\end{equation}
 where $\theta=\angle(\mathbf{r},\mathbf{r}^{\prime})$. Inserting
the latter expression into Eq.~(\ref{eq:A10}), and approximating
$|\mathbf{r}-\mathbf{r}^{\prime}|\simeq\mathbf{r}-\mathbf{r}\cdot\mathbf{r}^{\prime}/r$,
permits us to identify the wave vector at the point of observation
, $\mathbf{k}_{\textrm{out}}\equiv k\mathbf{r}/r$,

\begin{align}
\Psi_{\mathbf{k}}(\mathbf{r}) & =\phi_{\mathbf{k}}(\mathbf{r})-\frac{1}{4\gamma}\sqrt{\frac{i}{2k\pi r}}e^{ikr}\int d^{2}\mathbf{r}^{\prime}e^{-i\mathbf{k}_{\textrm{out}}\cdot\mathbf{r}^{\prime}}\times\nonumber \\
 & \times\left(1+\sigma_{2\theta(\mathbf{r},\mathbf{r}^{\prime})}\right)V_{d}(\mathbf{r}^{\prime})\Psi_{\mathbf{k}}(\mathbf{r^{\prime}}),\label{eq:BLG-6}\end{align}
 where the definition of $\sigma_{\theta}$ is given in Eq.~(\ref{eq:A15}).
As before, letting $\tilde{\sigma}_{2\theta}\equiv\sigma_{\theta(\mathbf{r},\mathbf{r}^{\prime}=0)}$,
and using Eq.~(\ref{eq:BLG-4}), we get\begin{align}
\Psi_{\mathbf{k}}(\mathbf{r}) & =\phi_{\mathbf{k}}(\mathbf{r})\nonumber \\
 & -\frac{2V_{0}}{8\gamma+iV_{0}}\sqrt{\frac{i}{2k\pi r}}e^{ikr}\left(1+\tilde{\sigma}_{2\theta}\right)u_{\mathbf{k}}^{(2)}\,.\label{eq:BLG-7}\end{align}
 The action of the last term on the spinor $u_{\mathbf{k}}^{(2)}$
yields the bilayer Berry phase term; taking the incident momentum
along the $x$ axis, $\mathbf{k}=(k,0)$, we obtain\begin{eqnarray}
\left(1+\tilde{\sigma}_{2\theta}\right)u_{\mathbf{k}}^{(2)} & = & \Xi_{B}(2\theta)\frac{1}{\sqrt{2}A}\left(\begin{array}{c}
1\\
e^{2i\theta}\end{array}\right)\nonumber \\
 & \equiv & \Xi_{B}(2\theta)u_{\mathbf{k}_{\textrm{out}}}^{(2)}\,,\label{eq:BLG-8}\end{eqnarray}
 where $\Xi_{B}$ is defined in Eq.~(\ref{eq:berry}) and $\theta$
is the scattering angle, $\theta=\angle\left(\mathbf{k},\mathbf{k}_{\textrm{out}}\right)$
{[}recall Eq.~(\ref{eq:A15}) and comments therein{]}. The wave function
of the scattered particles is then\begin{equation}
\Psi_{\mathbf{k}}(\mathbf{r})=\phi_{\mathbf{k}}(\mathbf{r})+f(\theta)\frac{e^{ikr}}{\sqrt{r}}u_{\mathbf{k}_{\textrm{out}}}^{(2)}\,,\label{eq:BLG-9}\end{equation}
 with the scattering amplitude reading\begin{equation}
f(\theta)=-\sqrt{\frac{i}{2k\pi}}\frac{2V_{0}}{8v_{F}^{2}\hbar^{2}/t_{\perp}+iV_{0}}\Xi_{B}(2\theta)\,,\label{eq:BLG-10}\end{equation}
 where we have restored all the constants. The FBA is recovered in
the limit $V_{0}\ll$energy scales,\begin{equation}
f_{\textrm{Born}}(\theta)=-\frac{V_{0}}{4v_{F}^{2}\hbar^{2}/t_{\perp}}\sqrt{\frac{i}{2k\pi}}\Xi_{B}(2\theta)\,.\label{eq:BLG-11}\end{equation}
 In contrast, in the limit of interest $V_{0}\rightarrow\infty$,
we obtain\begin{equation}
f_{V\rightarrow\infty}(\theta)=-\sqrt{\frac{2}{i\pi}}\frac{\Xi_{B}(2\theta)}{\sqrt{k}}\,.\label{eq:BLG-12}\end{equation}

\subsection*{A3. The dc-conductivity of monolayer and bilayer graphene}

The dc conductivity follows from the Boltzmann equation (see Sec.~\ref{sub:The-Boltzmann-approach}).
The expression for the semiclassical current $\bm{j}$ can be manipulated
to yield a more convenient form of the conductivity for our purposes.
We reproduce the main steps; at $T=0$ the Fermi function becomes
the Heaviside function, $\theta(\epsilon_{k_{F}}-\epsilon_{k})$,
and hence the expression for the current (including spin and valley
degeneracies) reads\begin{equation}
\bm{j}=\frac{g_{s}g_{v}e^{2}}{(2\pi)^{2}}\int d\mathbf{k}\,\tau(k)\delta(\epsilon_{k_{F}}-\epsilon_{k})(\mathbf{v}_{k}\cdot\mathbf{E})\mathbf{v}_{k}\,.\label{eq:aux1}\end{equation}
 Performing the angular integration, and using the relation $v_{r}=\hbar^{-1}\partial_{k}\epsilon$,
leads to\begin{equation}
\bm{j}=\frac{e^{2}}{\pi\hbar}\frac{k_{F}}{|v_{r}(k_{F})|}\tau(k_{F})(\mathbf{v}_{k_{F}}\cdot\mathbf{E})\mathbf{v}_{k_{F}}\,.\label{eq:aux}\end{equation}
 The longitudinal dc-conductivity follows from the latter expression:
\begin{equation}
\sigma_{\textrm{dc}}=\frac{2e^{2}}{h}\tau(k_{F})|v_{r}(k_{F})|k_{F}\,.\label{eq:sigma_DC_graph_and_bilayer}\end{equation}
 Using the results in Secs. A1 and A2 and the definition of relaxation
time $\tau(k_{F})$ (Sec.~\ref{sub:The-Boltzmann-approach}), we
can readily obtain the dc conductivity in the regime of $V_{0}\gg$energy
scales. (For a discussion of the on-site energy $V_{0}$ magnitude
see Sec.~\ref{sec_resonant}.)

The dc conductivity in the limit $V_{0}\rightarrow\infty$ reads\begin{equation}
\sigma_{\textrm{dc}}^{\textrm{strong}}=\begin{cases}
\frac{4e^{2}}{h}\frac{k_{F}^{2}}{2\pi^{2}n_{i}}\ln^{2}(k_{F}R) & \;\textrm{for monolayer}\\
\frac{4e^{2}}{h}\frac{k_{F}^{2}}{16n_{i}} & \;\textrm{for bilayer}\end{cases}\,.\label{eq:cond_rs}\end{equation}
 As expected, the dependence on $k_{F}$ coincides with that obtained
through the partial wave expansion method employed in Sec.~\ref{DCcond}.
Remarkably, the expressions match exactly {[}compare with Eq.~(\ref{eq_sigma_graphene})
and (\ref{eq_sigma_BL}){]}. This entails that scattering of a hard
disk of radius $R\sim a_{0}$ and scattering off a strong delta potential
have the same dependence on the momentum of the incident particles
(in both monolayer and bilayer graphene).

\bibliographystyle{apsrev4-1}

\begin{thebibliography}{77}
\bibitem{comment} A more accurate calculation of the density of states
starts from the spectrum as calculated directly from the four-band
model, Eq.~(\ref{eq_HBLfull}), and yields $\rho(E)=(t_{\perp}+2|E|)/(\pi\sqrt{3}t^{2})$
(for $|E|<t_{\perp}$) . The equation given in the text is recovered
for $|E|\ll t_{\perp}$.
\end{thebibliography}

\end{document}